\documentclass[12pt]{article}
\newcommand{\minus}[1]{ \mbox{$ (-1)^{#1} $}}

\newcommand{\CGC}[6]{ \mbox{$ \left( #1 #2 , #3 #4\,|\, #5 #6 \right) $} }

\newcommand{\SechsJ}[6]{ \mbox{$ %
            \arraycolsep0.25ex %
            \left\{ \begin{array}{ccc} %
                       #1 & #2 & #3 \vspace{0.5ex}\\%
                       #4 & #5 & #6 %
                   \end{array} \right\} $} }

\def\sumint{\hbox{\small$\Sigma$}\kern-0.73em\int\kern.1em}

\usepackage{authblk}
\usepackage{scicite}
\usepackage{times}
\usepackage{graphicx}
\usepackage{xcolor}
\usepackage{array,multirow,graphicx}
\usepackage{multirow,hhline,graphicx,array}
\usepackage{rotating}
\usepackage[superscript,biblabel]{cite}
\newcolumntype{M}[1]{>{\centering\arraybackslash}m{#1}}
\newcommand\RotText[1]{\rotatebox{90}{\parbox{2cm}{\centering#1}}}

\newenvironment{sciabstract}{%
\begin{quote} \bf}
{\end{quote}}

\title{Attosecond Pulse-shaping using a seeded Free Electron Laser}
\author[1]{Praveen Kumar Maroju}\author[2]{Cesare Grazioli}\author[3]{Michele Di Fraia}\author[1]{Matteo Moioli}\author[1]{Dominik Ertel}\author[1]{Hamed Ahmadi}
\author[3]{Oksana Plekan}\author[3]{Paola Finetti}\author[3]{Enrico Allaria}\author[3,4]{Luca Giannessi}\author[3,5]{Giovanni De Ninno}\author[3]{Carlo Spezzani}\author[3]{Giuseppe Penco}\author[3]{Alexander Demidovich}\author[3]{Miltcho Danailov}\author[3]{Roberto Borghes}\author[3]{Georgios Kourousias}\author[3]{Carlos Eduardo Sanches Dos Reis}\author[3]{Fulvio Bill\'e}\author[6]{Alberto A. Lutman}\author[7]{Richard J. Squibb}\author[7]{Raimund Feifel}\author[8]{Paolo Carpeggiani}\author[9]{Maurizio Reduzzi}\author[10]{Tommaso Mazza}\author[10]{Michael Meyer}
\author[11]{Samuel Bengtsson}\author[11]{Neven Ibrakovic}\author[11]{Emma Rose Simpson}\author[11]{Johan Mauritsson}
\author[12]{Tam\'{a}s Csizmadia}\author[12]{Mathieu Dumergue}\author[12]{Sergei K\"{u}hn}\author[12]{Harshitha N.G.}
\author[13]{Daehyun You}\author[13]{Kiyoshi Ueda}
\author[14]{Marie Labeye}\author[14]{Jens Egebjerg B\ae kh\o j}\author[14]{Kenneth J. Schafer}
\author[15]{Elena V. Gryzlova}\author[15]{Alexei N. Grum-Grzhimailo}
\author[3]{Kevin C. Prince}\author[3]{Carlo Callegari}\author[1]{Giuseppe Sansone}
\affil[1]{Physikalisches Institut, Albert-Ludwigs-Universit\"{a}t Freiburg Hermann-Herder-Stra{\ss}e 3, 79104 Freiburg, Germany.}
\affil[2]{ISM-CNR, Trieste LD2 Unit, Basovizza AREA Science Park, Trieste, I-34149, Italy}
\affil[3]{Elettra-Sincrotrone Trieste, 34149 Basovizza, Trieste, Italy.}
\affil[4]{ENEA C.R. Frascati, Via E. Fermi 45, 00044 Frascati (Roma).}
\affil[5]{Laboratory of Quantum Optics, University of Nova Gorica, 5001 Nova Gorica, Slovenia.}
\affil[6]{SLAC National Accelerator Laboratory, Menlo Park, California 94025, USA}
\affil[7]{Department of Physics, University of Gothenburg, Origov\"{a}gen 6B, 412 96 Gothenburg, Sweden.}
\affil[8]{Technische Universit\"{a}t Wien, Austria.}
\affil[9]{Dipartimento Fisica Politecnico, Piazza Leonardo da Vinci 32, 20133 Milano Italy.}
\affil[10]{European XFEL, Holzkoppel 4, 22869 Schenefeld, Germany.}
\affil[11]{Department of Physics, Lund University, PO Box 118, SE-221 00 Lund, Sweden.}
\affil[12]{ELI-ALPS, ELI-Hu Kft., Dugonics t\'{e}r 13, H-6720 Szeged Hungary.}
\affil[13]{Institute of Multidisciplinary Research for Advanced Materials, Tohoku University, Sendai 980-8577, Japan.}
\affil[14]{Department of Physics and Astronomy Louisiana State University Baton Rouge, Louisiana 70803-4001, USA}
\affil[15]{Skobeltsyn Institute of Nuclear Physics, Lomonosov Moscow State University, Moscow 119911, Russia.}

\date{}

\begin{document}


\baselineskip24pt


\maketitle


\begin{sciabstract}
\begin{sciabstract}
Attosecond pulses are fundamental for the investigation of valence and core-electron dynamics on their natural timescale~\cite{RMP-Krausz-2009,NATPHYS-Corkum-2007,SCIENCE-Kapteyn-2007}.
At present the reproducible generation and characterisation of attosecond waveforms has been demonstrated only through the process of high-order harmonic generation~\cite{Paul2001b, NATURE-Kienberger-2004, NATURE-Tzallas-2003, PRL-Nabekawa-2006}.
Several methods for the shaping of attosecond waveforms have been proposed, including metallic filters~\cite{PRL-Lopez-2005, OL-Gustafsson-2007},
multilayer mirrors~\cite{OL-Hofstetter-2011} and manipulation of the driving field~\cite{NATURE-Bartels-2000}.
However, none of these approaches allow for the flexible manipulation of the temporal characteristics of the attosecond waveforms,
and they suffer from the low conversion efficiency of the high-order harmonic generation process.
Free Electron Lasers, on the contrary, deliver femtosecond, extreme ultraviolet and X-ray pulses with energies ranging from tens of $\mathrm{\mu}$J to a few mJ~\cite{NATPHOT-Ackermann-2007,NATPHOT-Emma-2010}.
Recent experiments have shown that they can generate sub-fs spikes, but with temporal characteristics
that change shot-to-shot~\cite{APL-Marinelli-2017, PRL-Huang-2017, NATPHOT-Hartmann-2018}.
Here we show the first demonstration of reproducible generation of high energy ($\mathrm{\mu}$J level) attosecond waveforms using a
seeded Free Electron Laser~\cite{NATPHOT-Allaria-2012}.
We demonstrate amplitude and phase manipulation of the harmonic components of an attosecond pulse train in combination with a novel approach
for its temporal reconstruction.
The results presented here open the way to perform attosecond time-resolved experiments with Free Electron Lasers.

\end{sciabstract}

\end{sciabstract}

The intensities and the relative phases between the harmonics $q\omega_F$ (with $q$ integer and $\omega_F$ fundamental frequency)
in an extreme ultraviolet (XUV) comb determine the temporal structure of the resulting attosecond pulse train.
The intensities of the harmonics can be easily measured using a (photon or electron) spectrometer.
Phase information, which is harder to come by, is usually obtained by observing the interference between different pathways leading
to states with the same final energy, where the phase to be characterised is included in at least one of the pathways. With XUV pulses,
the natural observable is a photoelectron, hence different pathways into the ionisation continuum are studied.
The XUV frequency comb produced by high-order harmonic generation (HHG) consists of odd-integer harmonics of the fundamental field,
and the ionisation process takes place in the presence of a near-infrared (NIR), dressing field with same frequency $\omega_F$.
Under these conditions, additional photons may be absorbed or emitted producing a single sideband halfway between the main photoelectron peaks.
Each sideband can be populated through two pathways leading to final states of the same parity and this results in a variation in sideband amplitude
as a function of the relative phase of the two pathways.  If the XUV and fundamental fields are precisely synchronised, as they can be in HHG,
then delaying the fields with respect to each other reveals the phase information~\cite{Paul2001b, Science-Mairesse-2003}.

\begin{figure}
\centering \resizebox{1.0\hsize}{!}{\includegraphics{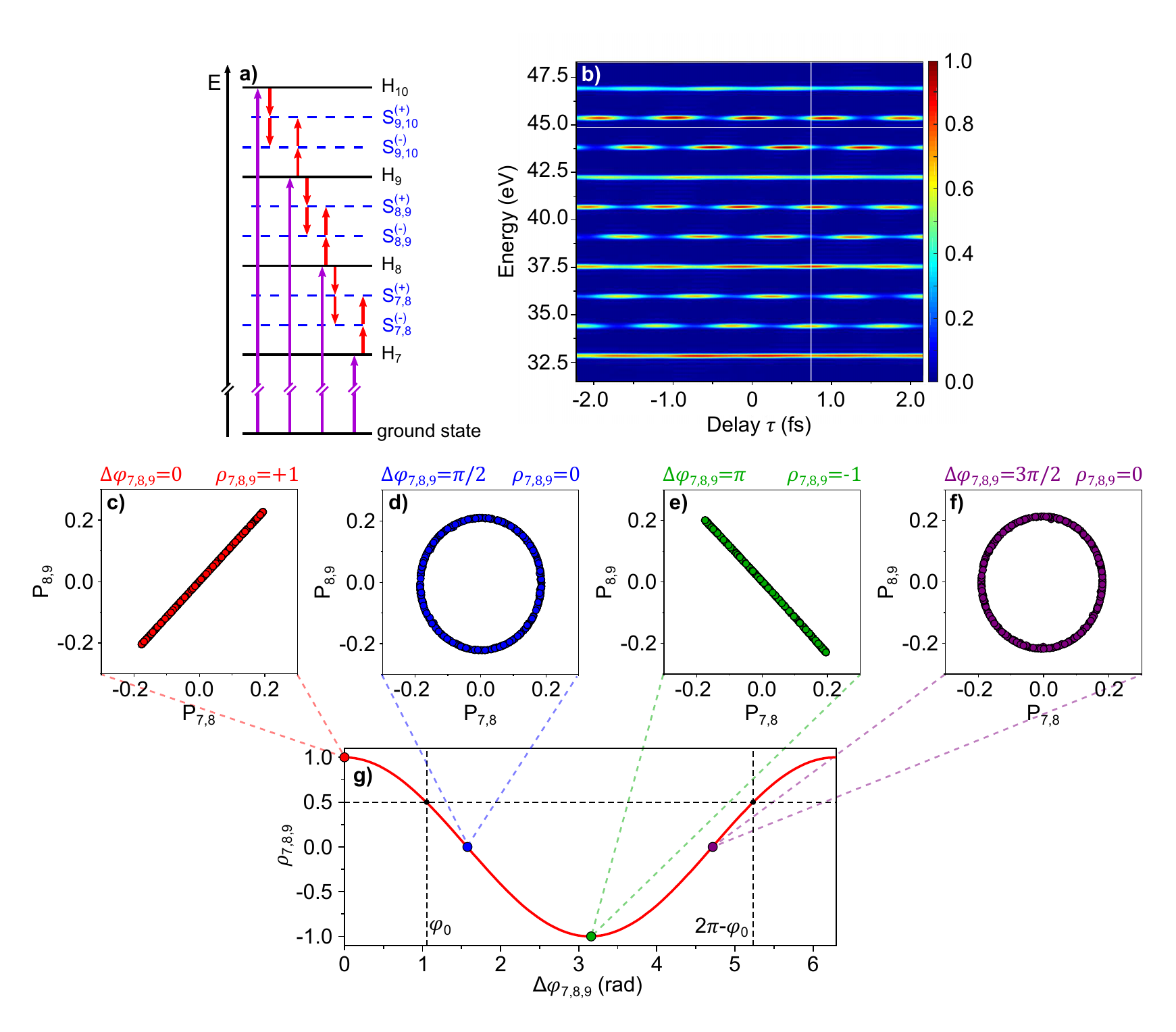}}
\caption{\textbf{Multi-photon sideband generation and principle of the measurement.} a) Schematic view of multi-NIR-photon sideband generation.
Energy levels of the photoelectrons generated by the harmonics of the FEL ($\mathrm{H_7}-\mathrm{H_{10}}$; magenta arrows) and by the additional absorption
and emission of one and two NIR photons ($S^{(\pm)}_{q,q+1}$; red arrows). b) Expected photoelectron spectra as a function of the relative delay
$\tau$ between the train of attosecond pulses and the NIR field along the (positive) common direction of polarisation of the two fields.
The photoelectron spectra are characterised by an oscillation with a period $T=2\pi/(3\omega_{NIR})$.
Correlation plots of the oscillating components of the sidebands (c, d, e, f) for four phase differences $\Delta\varphi_{7,8,9}$: $0$ (c), $\pi/2$ (d), $\pi$ (e), $3\pi/2$ (f).
g) Evolution of the correlation coefficient $\rho_{7,8,9}$ as a function of $\Delta\varphi_{7,8,9}$. The intensity of the NIR pulse is $I_{NIR}=1.5\times10^{11}~\mathrm{W/cm^2}$.}
\label{Fig1}
\end{figure}

In our study, the harmonic comb was generated by the seeded Free Electron Laser (FEL) FERMI, which uses an ultraviolet pulse ($\omega_{UV}=\omega_F=4.69$~eV)
derived from a frequency-tripled NIR pulse ($\omega_{NIR}=\omega_{UV}/3$) as seed.
Three ($q=7,8,9$) and four ($q=7,8,9,10$) harmonics of $\omega_{UV}$ were generated using two different undulator configurations
(see Fig.~\ref{Fig1n_ED} and Table~\ref{Table1n}). To characterise the pulses, photoionisation took place in the presence of a field with frequency
$\omega_{NIR}$ leading to the formation of  two sidebands between each pair of the main XUV peaks (see Fig.~\ref{Fig1}a).
The two sidebands $S^{(\pm)}_{q,q+1}$ can be each populated through two paths characterised by a different number of exchanged NIR photons: the absorption of one photon of the harmonic $q$ and one (two) NIR photon, or the absorption of one photon of the harmonic $q+1$ and the emission of two (one) NIR photons. The difference in parity for the final states of the interfering pathways determines an asymmetry in the photoionisation emission. If the observation is restricted along a single direction, the intensity of the sidebands oscillates as a function of the delay $\tau$ between the NIR and XUV pulse (see Fig.~\ref{Fig1}b):

\begin{equation}\label{sideband parameter}
S^{(\pm)}_{q,q+1}(\tau)\propto 1\pm\alpha_{q,q+1}\cos\left[\varphi_{q+1}-\varphi_{q}+3\omega_{NIR}\tau\right]=1\pm P_{q,q+1}(\tau),
\end{equation}
where $\alpha_{q,q+1}$ depends on the intensity and
energy of the two harmonics $q$ and $q+1$ with phases $\varphi_{q}$ and $\varphi_{q+1}$, on the photoelectron energy,
and on the intensity of the NIR pulse, and the equality defines the oscillating component of the sideband intensity $P_{q,q+1}$ under the approximations detailed in the Supplementary Information (SI). If the delay $\tau$ could be precisely controlled,
then the relative phase between consecutive harmonics could be estimated from the time shift between the oscillations of the sidebands.

This approach cannot be applied directly for the reconstruction of the relative phase of multiple harmonics generated by a FEL due to the lack of
sub-cycle synchronisation between the harmonics and the NIR field~\cite{NATCOMM-Schulz-2015}, which completely washes out the delay-dependence of
the sideband oscillations. The information can be still retrieved, however, through a correlation analysis of the fluctuating sideband intensities
measured on a single-shot basis.

This novel approach is presented in Fig.~\ref{Fig1}, c, d, e, and f, which show the simulated correlation plots of the oscillating
components $P_{8,9}$ and $P_{7,8}$ of the sidebands for a random variation of the delay $\tau$ in the range +/-3~fs, which is the typical delay jitter
measured in the experiment~\cite{OE-Danailov-2014} .
The correlation plot eliminates the explicit dependence on $\tau$ and results in an ellipse, whose shape depends on the phase difference:
\begin{equation}\label{Eq5}
\Delta\varphi_{q-1,q,q+1}=\varphi_{q+1}+\varphi_{q-1}-2\varphi_{q}.
\end{equation}
The intensity profile of the pulse train depends only on this phase difference (apart from a trivial time shift; see SI).
Depending on the phase difference $\Delta\varphi_{7,8,9}$, the plot evolves from a linear distribution with positive correlation (Fig.~\ref{Fig1}c),
to a circle (Fig.~\ref{Fig1}d), to a linear distribution with negative correlation (Fig.~\ref{Fig1}e), and finally back to a circle (Fig.~\ref{Fig1}f).
These changes clearly indicate that the shape of the correlated distribution is related to the synchronisation of the three harmonics
(the complete evolution as a function of the phase difference is presented in Fig.~\ref{Fig4n_ED}).
The phase information can be derived from the distribution by evaluating its correlation coefficient $\rho_{q-1,q,q+1}$ (see SI),
which quantifies the extent to which the two sidebands oscillate perfectly in- ($\Delta\varphi_{q-1,q,q+1}=0$ and $\rho_{q-1,q,q+1}=+1$) or
out-of-phase ($\Delta\varphi_{q-1,q,q+1}=\pi$ and $\rho_{q-1,q,q+1}=-1$).
The correlation coefficient oscillates as a function of the phase difference $\Delta\varphi_{7,8,9}$ as shown Fig.~\ref{Fig1}g,
and it closely resembles a cosine function.
Two different values of $\Delta\varphi_{7,8,9}$ correspond to the same value of the correlation coefficient: $\varphi_0$ and $2\pi-\varphi_0$.
This ambiguity can be resolved in the experiment by controlling the modulus and sign of the phase differences between the harmonics (see SI).
Simulations based on the solution of the time-dependent Schr\"odinger equation confirmed the validity of our approach (see Fig.~\ref{Fig5n_ED}).

In the experiment, the intensity of the harmonics was independently controlled by tuning the undulator gaps and the dispersive section of the electron
transport optics. The phase between the harmonics was controlled by phase shifters~\cite{NATPHOT-Prince-2016, PRL-Iablonskyi-2017},
which introduce a delay $\tau_{si}$ ($i$ indicates the $i$th-phase shifter) for a selected harmonic $q$, affecting the phase difference $\Delta\varphi_{q-1,q,q+1}$,
through a term $q\omega_{UV}\tau_{si}$ (see Fig.~\ref{Fig1n_ED}). In this respect, seeded FELs offer a superior degree of control with respect to HHG-sources,
for which the intensities and phases of the single harmonic cannot be independently controlled.\\
\begin{figure}
\centering \resizebox{1.0\hsize}{!}{\includegraphics{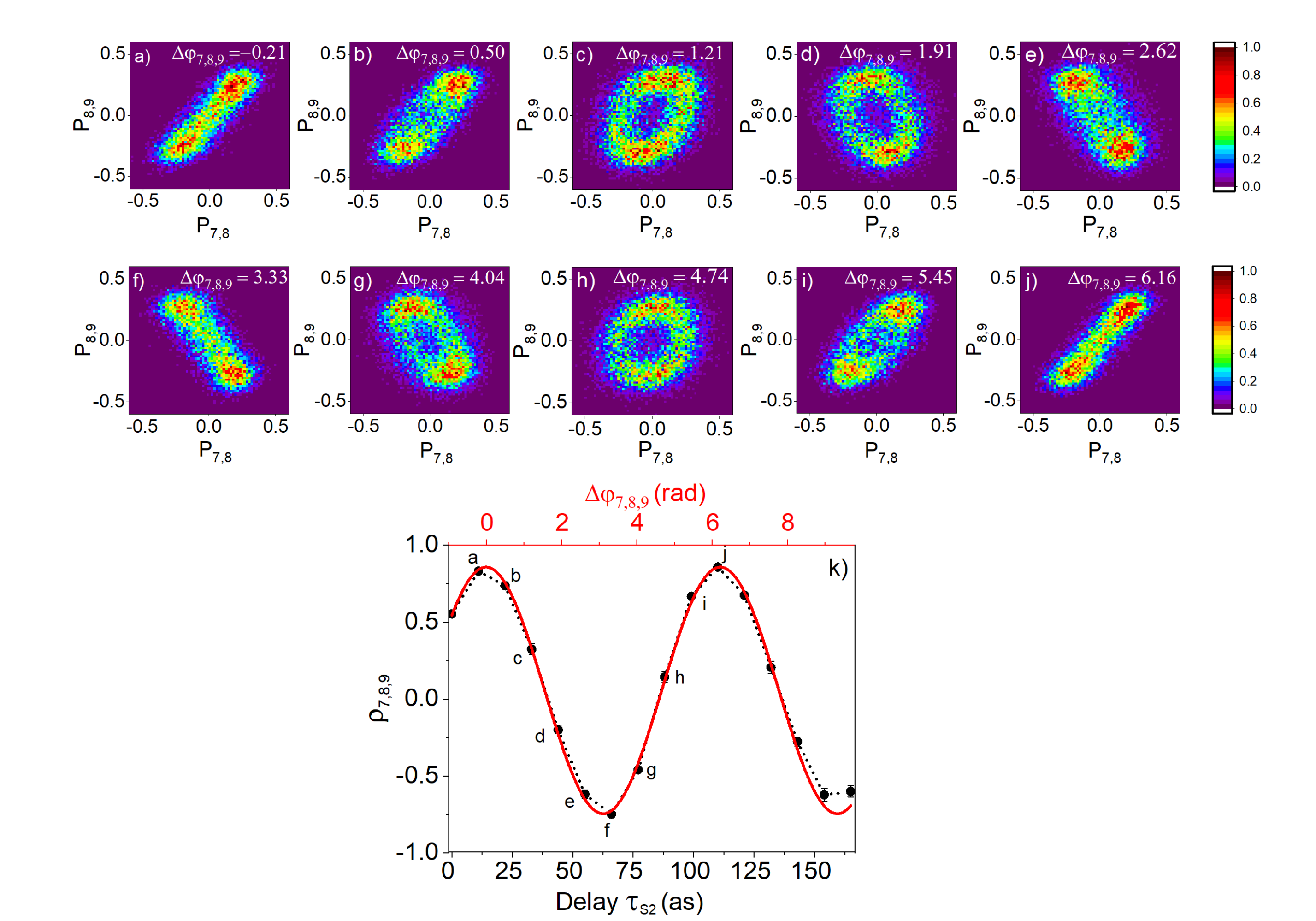}}
\caption{\textbf{Correlation plots of the oscillating components of the sidebands and phase difference $\Delta\varphi_{7,8,9}$ retrieval.}
Evolution of the correlation plots of the oscillating components of the sidebands for increasing values of the delay $\tau_{s2}$ introduced by the phase shifter $\mathrm{PS_2}$ (a-j) (see Fig.~\ref{Fig1n_ED}). Evolution of the correlation coefficient $\rho_{7,8,9}$ as a function of the delay $\tau_{s2}$ (black points and dotted line) and sinusoidal fit (red) (k). The $\Delta\varphi_{7,8,9}$ upper x-axis (red) was obtained by assigning the maxima of the fit to the values $\Delta\varphi_{7,8,9}=2m\pi$, where $m$ is an integer. The intensity of the NIR pulse was estimated to be $I_{NIR}\approx1.5\times 10^{12}~\mathrm{W/cm^2}$. The value of the correlation parameters, the phase differences and the corresponding errors are presented in Tab.~\ref{Table2n}}
\label{Fig2}
\end{figure}
Figure~\ref{Fig2}a-j  presents the experimental results for the three-harmonic case, for different delays $\tau_{s2}$ introduced on the 9th harmonic. 
These measurements indicate a periodic evolution of the correlated distributions in close agreement with the theoretical prediction.
A partial broadening of the distributions is attributed to the shot-to-shot fluctuations of the single harmonic intensity.
The correlation coefficients $\rho_{7,8,9}$ (black points and dotted line in Fig.~\ref{Fig2}k) and the fit (red curve) clearly follow a cosine
evolution in good agreement with the simulations.
The maxima of the fit were assigned to the phase differences $\Delta\varphi_{7,8,9}=0,2\pi$ (see upper x-axis in Fig.~\ref{Fig2}k)
and the curve was used to assign a phase difference $\Delta\varphi_{7,8,9}$ to each delay $\tau_{s2}$. \textcolor{black}{The error in the estimation of the phase difference
depends on the slope of the curve (which depends on the NIR intensity) and it was typically in the range 0.05-0.1 rad for our experimental conditions (see SI)}.
\textcolor{black}{The characterisation of pulses with reproducible temporal structure gives the possibility to accumulate data over several single-shot measurements,
thus improving the signal-to-noise ratio and reducing the error in the temporal reconstruction.
In the case of a self-amplified spontaneous emission FEL, pulse properties change on a shot-to-shot basis and a single-shot technique is therefore mandatory~\cite{NATPHOT-Hartmann-2018}.}

\begin{figure}
\centering \resizebox{1.0\hsize}{!}{\includegraphics{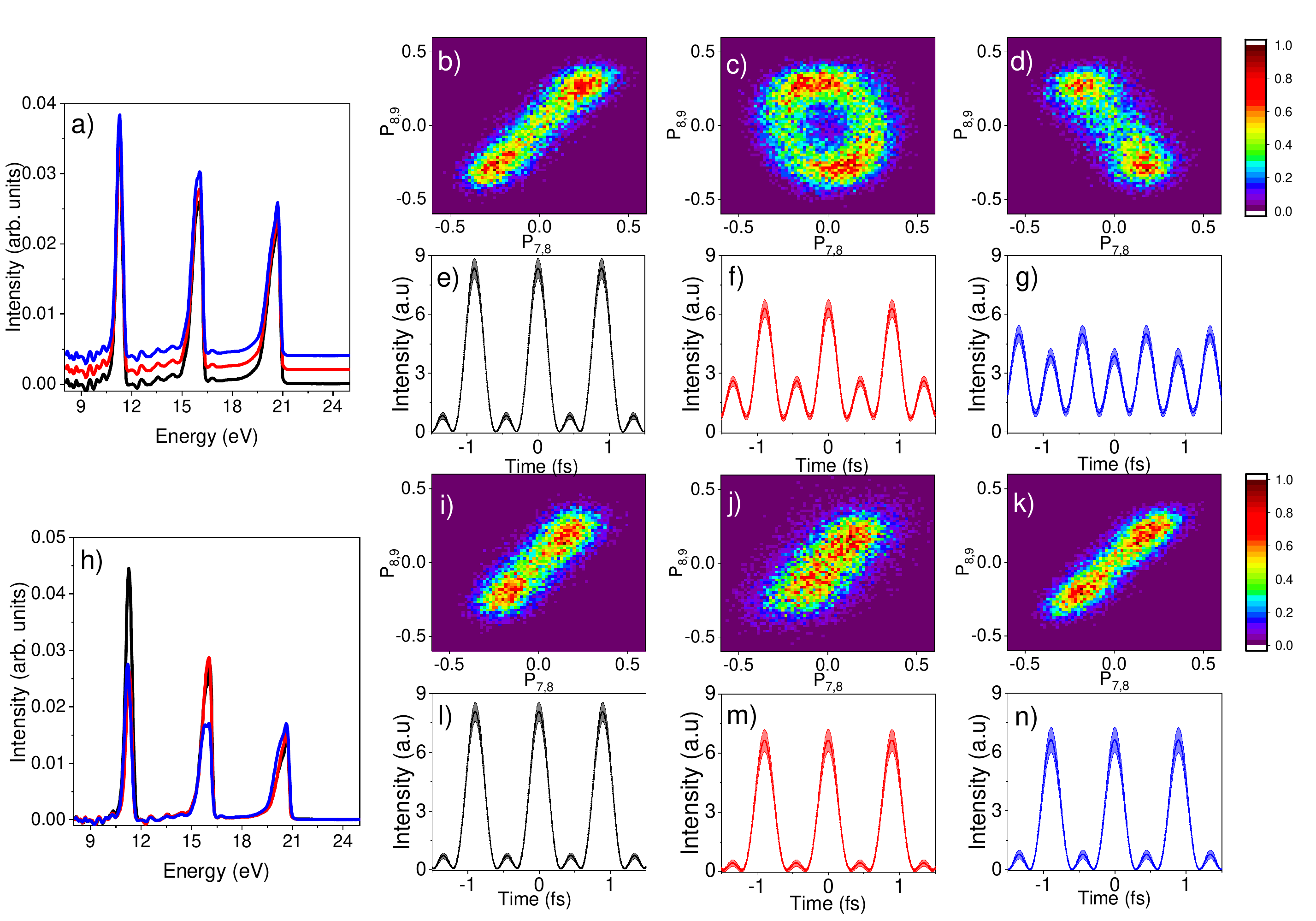}}
\caption{\textbf{Complete phase and amplitude shaping of attosecond waveforms.} Photoelectron spectra (a), correlation plots of the oscillating components
of the sidebands (b, c, d) and reconstructed attosecond waveforms (e, f, g)
in the case of independent phase shaping for three phase differences $\Delta\varphi_{7,8,9}$ (black curve and panels b, e;
red curve and panels c, f; blue curve and panels d, g).
The three photoelectron spectra in (a) were vertically shifted for visual clarity.
Photoelectron spectra (h), correlation plots of the oscillating components of the sidebands (i, j, k)
and reconstructed attosecond waveforms (l, m, n) in the case of independent amplitude control for three settings
of the harmonic amplitudes (black curve and panels i, l; red curve and panels j, m; blue curve and panels k, n)
using the same values of the phase shifters. See Table~\ref{Table3n} for additional information on the phase difference $\Delta\varphi_{7,8,9}$
and the amplitude of the harmonics $F_7, F_8,$ and $F_9$. \textcolor{black}{The errors in the reconstruction of the attosecond pulse trains are determined by the error bars for the amplitude and phase differences
(see Table~\ref{Table3n}) and are indicated as shaded in (e, f, g, l, m, n).}}
\label{Fig3}
\end{figure}

We exploited this novel approach for the determination of the relative phase of XUV harmonics to demonstrate the independent phase-amplitude shaping capability of attosecond waveforms offered by the FEL FERMI. 
Figure~\ref{Fig3}a and Fig.~\ref{Fig3}, b, c and d show three photoelectron spectra and the corresponding correlation plots for three phase differences $\Delta\varphi_{7,8,9}$, respectively. The phase change does not appreciably modify the intensities of the three harmonics (for the amplitudes $F_j$ of the single $j$th harmonic and for the phase differences see Table~\ref{Table3n}).
The reconstructed intensity profiles $I(t)$ are presented in Fig.~\ref{Fig3}, e, f and g. The measurements indicate a pure phase shaping of the harmonic comb: a well-defined attosecond pulse train (Fig.~\ref{Fig3}e) obtained for $\Delta\varphi_{7,8,9}=0.08\pm0.08$ rad (close to the ideal condition of harmonics in phase $\Delta\varphi_{7,8,9}=0$) is transformed first into an attosecond pulse train of lower amplitude with a satellite (Fig.~\ref{Fig3}f), when $\Delta\varphi_{7,8,9}=1.93\pm0.03$, and finally into an attosecond pulse train characterised by a double structure for $\Delta\varphi_{7,8,9}=3.29\pm0.24$ (Fig.~\ref{Fig3}g),
which is close to the condition of harmonics out-of-phase $\Delta\varphi_{7,8,9}=\pi$.
Figure~\ref{Fig3}h shows three photoelectron spectra corresponding to three different settings of the amplitudes of the three harmonics (see Table~\ref{Table3n}).
The amplitude of the single harmonic was modified by about $25\%$ using the dispersive section and the undulator gaps (see SI).
Figure~\ref{Fig3}, i, j and k show the correlation plots for the same position of the phase shifters. The phase difference $\Delta\varphi_{7,8,9}$ remains constant within the experimental error, independent from the variations of the single harmonic intensity. 
The reconstructed attosecond pulse trains for the three configurations are presented in Fig.~\ref{Fig3}, l, m and n. These data demonstrate a pure amplitude shaping of the harmonic comb: the well-defined attosecond pulse structure ($\Delta\varphi_{7,8,9}$ is close to zero ($2\pi$) for the three measurements) is preserved for the three configurations and the different harmonic intensities lead only to a variation in the maxima of the intensity profiles.
\textcolor{black}{The energy of the attosecond pulse train was about $16~\mu$J. Table-top-based HHG source yield much lower energies (in the nJ range),
and only few experimental groups have reported total pulse energies on target approaching the $\mu$J range~\cite{NATCOMM-Takahashi-2013, PRA-Nayak-2018}}.

\textcolor{black}{We estimated that a pulse energy of about 50~nJ per harmonic is sufficient for the acquisition of single-shot photoelectron spectra,
which is well below the typical energy per harmonic (a few $\mu$J) available at FERMI.}
\textcolor{black}{The currently available range of seed wavelengths at FERMI (360—230 nm) would allow a moderate control of the comb periodicity around 1~fs,
however a revised layout of the seed laser optimized for this task could increase the spike separation to tens of fs.
Shorter separation can be already achieved by using alternate harmonics (e.g., $q = 6, 8, 10$)}

\textcolor{black}{We should point out that alternative FEL-based approaches have been theoretically proposed for the generation of a train of
attosecond pulses~\cite{PRSTAB-Zholents-2005, PRL-Thompson-2008}. Even though the predicted peak power levels (GW) and pulse
durations (down to the sub-100 attoseconds) are comparable with those reported here, or in principle achievable with our approach, these methods do not offer a strategy for controlling the relative amplitudes and phases of the single harmonics, i.e. for attosecond pulse shaping.}
\textcolor{black}{Extension of our approach to wavelength as short as 4~nm (300~eV) appear feasible if combined with
the Echo-enabled harmonic generation (EEHG) seeding scheme \cite{NATPHOT-Rebernik-2019}. Numerical simulations indicate
that the method for the temporal characterization could be applied for photon energies from $\sim$20~eV up to 300~eV,
using a suitable gas target.}

\begin{figure}
\centering \resizebox{1.0\hsize}{!}{\includegraphics{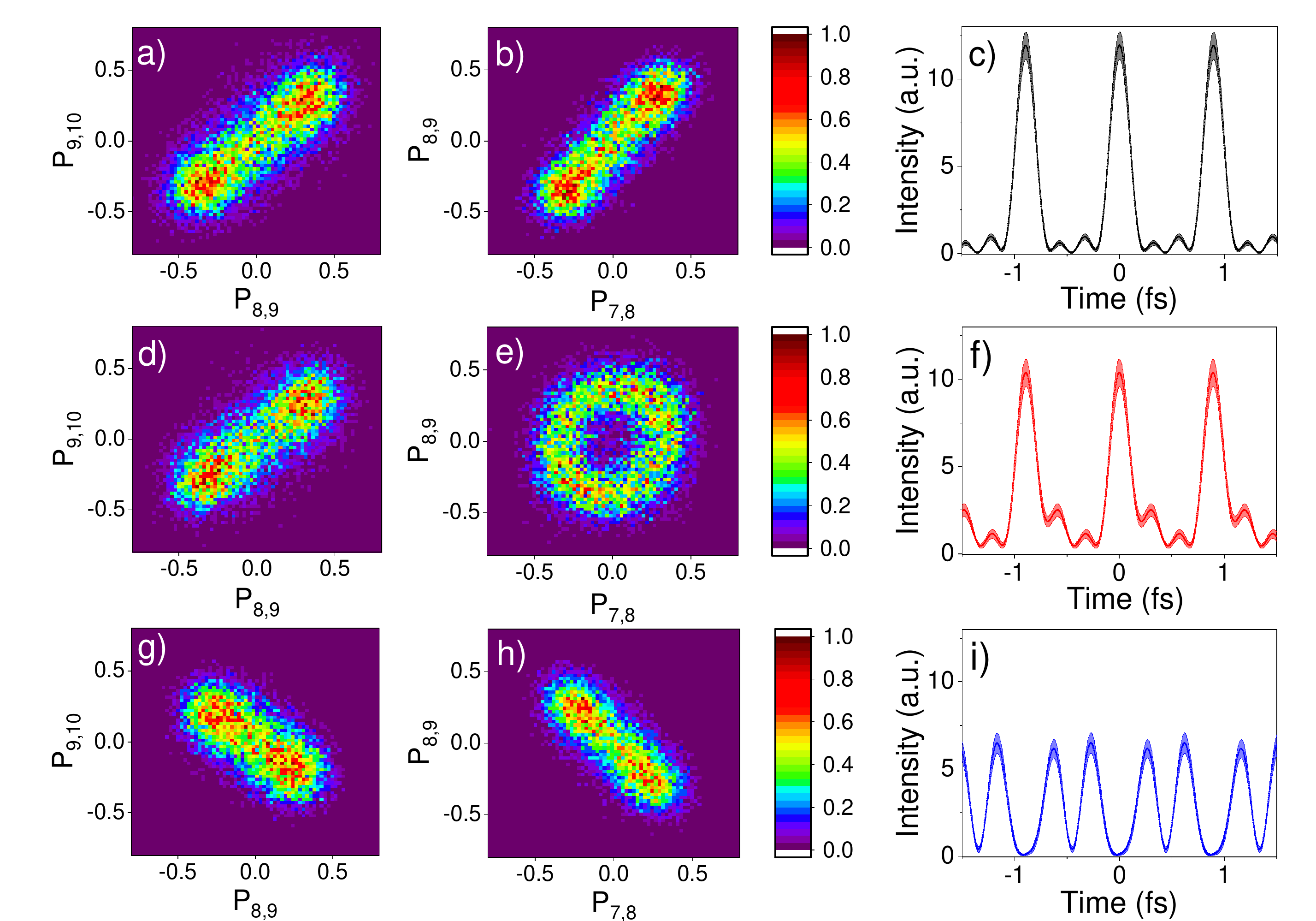}}
\caption{\textcolor{black}{\textbf{Synthesis of complex attosecond waveforms.}} Correlation plots of the oscillating components of the sidebands for
$\Delta\varphi_{8,9,10}$ (a, d, g) and $\Delta\varphi_{7,8,9}$ (b, e, h) and the retrieved attosecond waveforms (c, f, i)
for the four harmonic experiment and three different combinations of the phase differences $\Delta\varphi_{7,8,9}$
and $\Delta\varphi_{8,9,10}$ (a-c; d-f; g-i). See Table~\ref{Table3n} for additional information on the phase difference
$\Delta\varphi_{7,8,9}$ and $\Delta\varphi_{8,9,10}$, and the amplitude of the harmonics $F_7, F_8, F_9$, and $F_{10}$.
\textcolor{black}{The errors in the reconstruction of the attosecond pulse trains are determined by the error bars for the amplitude and phase differences
(see Table~\ref{Table3n}) and are indicated as shaded in (c, f, i).}}
\label{Fig4}
\end{figure}

As a first demonstration of complex attosecond waveform shaping, we considered the case of four harmonics (see Fig.~\ref{Fig4}), for which the attosecond temporal
structure depends on the two phase differences $\Delta\varphi_{7,8,9}$ and $\Delta\varphi_{8,9,10}$.
\textcolor{black}{The photoelectron spectra with (red curve) and without NIR (black curve) are shown in Fig.~\ref{Fig1n_ED} d}.
The independent control of the two phases gives the opportunity to generate ultrashort (chirp free) attosecond pulse trains,
as shown in  Fig.~\ref{Fig4}, a, b and c, which report the correlation plots $(P_{8,9}-P_{9,10})$ (a) and $(P_{7,8}-P_{8,9})$ (b)
in the case of maximum positive correlation (see Table~\ref{Table3n} for the values of the amplitude and phase differences).
\textcolor{black}{The reconstruction (Fig.~\ref{Fig4}c) returns a duration (FWHM) of the single pulse of about $210\pm4$~as.}
Figure~\ref{Fig4}, d and e present the results corresponding to $\Delta\varphi_{8,9,10}=0.20\pm0.15$ (close to the configuration of harmonics in phase $\Delta\varphi_{8,9,10}=0$)
and $\Delta\varphi_{7,8,9}=1.23\pm0.06$~rad (harmonics only partially in phase). \textcolor{black}{In the temporal domain, this condition translates into a partial broadening of the peaks (FWHM=$220\pm5$~as)
and the appearance of small satellites in the reconstructed attosecond pulse train (Fig.~\ref{Fig4}f).}
Finally, Fig.~\ref{Fig4}, g and h present the results when $\Delta\varphi_{8,9,10}=2.89\pm0.08$ rad and $\Delta\varphi_{7,8,9}=2.95\pm0.09$ (i.e. both phase differences are close to $\pi$), respectively:
the four harmonics are divided in two groups (harmonics 7-8, and harmonics 9-10), each pair of which is (approximately) in phase,
with an additional phase jump of $\pi$ between the two groups.
This condition leads to a double attosecond pulse structure, which is visible in the reconstruction presented in Fig.~\ref{Fig4}i.
\textcolor{black}{The availability of six undulators at FERMI would in principle allow for the generation of six harmonics. This configuration, however, may require a revised and optimized setup.
Simulations indicate that the experimental technique demonstrated in this work could be extended for the characterization of pulses with durations in the sub-100~as regime.}

Our technique also offers the possibility to determine with a sub-cycle resolution the relative phase of the XUV and NIR pulses, enabling phase-resolved pump-probe
experiments at FELs based on attosecond pulse trains (see SI and Fig.~\ref{Fig3n_ED}). The high intensities in the XUV and X-ray spectral region reached by FELs,
\textcolor{black}{combined with} the capabilities \textcolor{black}{offered by seeding to} independently control and shape the amplitudes and phases of
attosecond pulses will open new possibilities for the investigation and control of ultrafast nonlinear electronic processes.
\textcolor{black}{The ongoing design of future seeded FEL sources at other facilities such as LCLS~\cite{LCLS}, FLASH~\cite{FLASH}, SINAP~\cite{SINAP}
could be optimized in order to improve capabilities toward this method with respect to existing facilities like FERMI~\cite{NATPHOT-Allaria-2012} and
 DALIAN~\cite{DALIAN}.} In solid samples, attosecond shaped waveforms could be used to promote electrons from the inner-valence to the conduction band,
 giving the possibility to investigate diffusion and relaxation effects with attosecond resolution and with temporally sculpted electronic wave packets.
 More generally our results give access to programmable attosecond waveforms at high intensities for the first time.

\subsubsection*{Acknowledgments}

This project has received funding from the European Union's Horizon 2020 research and innovation programme under the Marie Sklodowska-Curie grant agreement no.~641789 MEDEA and the Italian Ministry of Research (Project FIRB No. RBID08CRXK). K.U. acknowledges support by the X-ray Free Electron Laser Utilization Research Project and the X-ray Free Electron Laser Priority Strategy Program of the Ministry of Education, Culture, Sports, Science and Technology of Japan (MEXT), by the Cooperative Research Program of ‘‘Network Joint Research Center for Materials and Devices: Dynamic Alliance for Open Innovation Bridging Human, Environment and Materials’’, by the bilateral project CNR-JSPS "Ultrafast science with extreme ultraviolet Free Electron Lasers", and by the IMRAM project for the international co-operation. R.F. and J.M. thank the Swedish Research Council (VR) and the Knut and Alice Wallenberg Foundation for financial support. E.V.G. acknowledges support of the Foundation for the Advancement of Theoretical Physics and Mathematics “BASIS”. Research at Louisiana State University was supported by the US Department of Energy, Office of Science, Basic Energy Sciences, under contract no. de-sc0010431. Portions of this research were conducted with high performance computing resources provided by Louisiana State University (http://www.hpc.lsu.edu) and Louisiana Optical Network Infrastructure (http://hpc.loni.org). G.S. acknowledges useful discussion about the simulations and the data analysis with Thomas Pfeifer and Matthias K\"ubel. We acknowledge L. Foglia, A. Simoncig, and M. Coreno for valuable discussions.

\subsubsection*{Authors contributions}
P.K.M., Ma.Mo., D.E., M.D.F., O.P., T.M., M.M., S.B., N.I., E.R.S., J.M., T.C., M.D., S.K., H.N.G., D.Y., K.U., K.P., C.G., and C.C.
contributed to the data acquisition and to data analysis. L.G., E.A. G.D.N, C.S., G.P. operated the machine and designed the three and four harmonic generation
scheme. A.L. contributed to the machine operation. A.D. and Mi.Da. designed the beam path for the NIR pulse. P.F. designed the mechanics for the recombination mirror.
A.D. and C.G. designed the recombination mirror, and the whole set-up was installed by A.D., C.G., and M.D.F.. A.D. and M.D.F. designed and installed the beam dump
diagnostic system for the alignment of collinear configuration. M.D.F., O.P., C.C. prepared the end station. R.B., C.K., C.E.S.D.R, F.B.
developed the analysis tools used during the beamtime. M.R. contributed to the preliminary development of the simulation codes. M.L., J.E.B.
and K.J.S. performed the TDSE calculations. R.J.S. and R.F. constructed and operated the magnetic bottle electron spectrometer,
where parts of its operation mode were conceptually devised for this experiment. A.N.G.-G. and E.V.G. developed the perturbation-theory approach and
derived the atomic phases contributions. C.C. and G.S. conceived the idea of the experiment.
G.S. developed the numerical code for the SFA simulations. P.K.M. developed the numerical code for the correlation analysis. P.K.M., C.C. and G.S.
analysed the experimental data and performed the simulations. G.S. supervised the work. P.K.M., A.N.G.-G., E.V.G., C.C. and G.S. wrote the manuscript,
which was discussed and agreed by all coauthors.\\

\subsubsection*{Author information}
Reprints and permissions information is available at www.nature.com/reprints.\\
The authors declare no competing interests.\\
\textcolor{black}{Raw data were generated at the FERMI large-scale facility.
Derived data supporting the findings of this study are available from the corresponding author upon request.}\\
Correspondence and requests for materials should be addressed to giuseppe.sansone@physik.uni-freiburg.de\\

\newpage

\section*{Methods}
\setcounter{figure}{0}
\setcounter{table}{0}
\renewcommand{\thefigure}{\arabic{figure} Extended Data}
\renewcommand{\thetable}{\arabic{table} Extended Data}
\subsection*{Experimental setup}
\begin{figure}[hb]
\centering \resizebox{1.0\hsize}{!}{\includegraphics{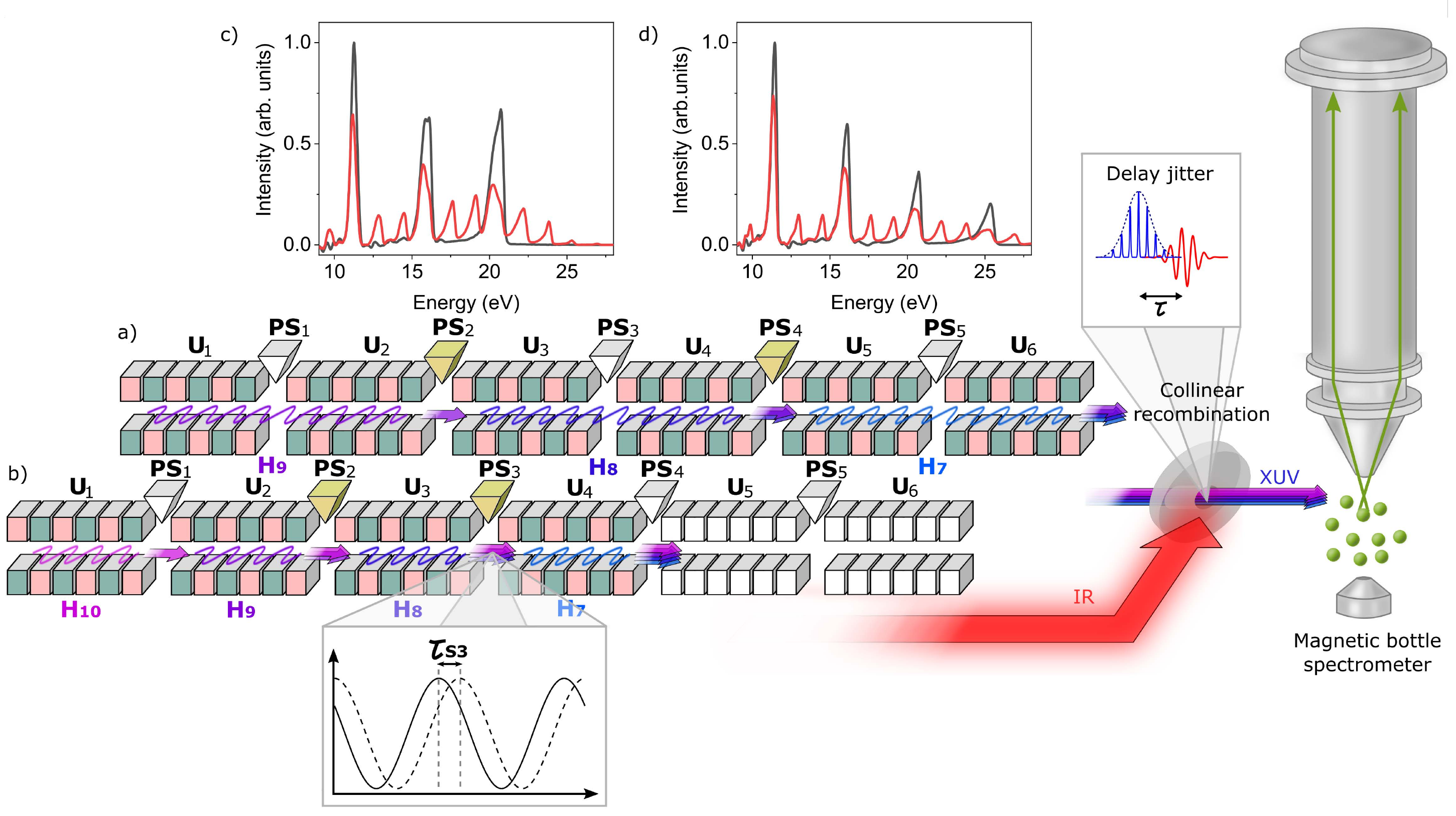}}
\centering \resizebox{0.8\hsize}{!}{\includegraphics[width=0.4\textwidth]{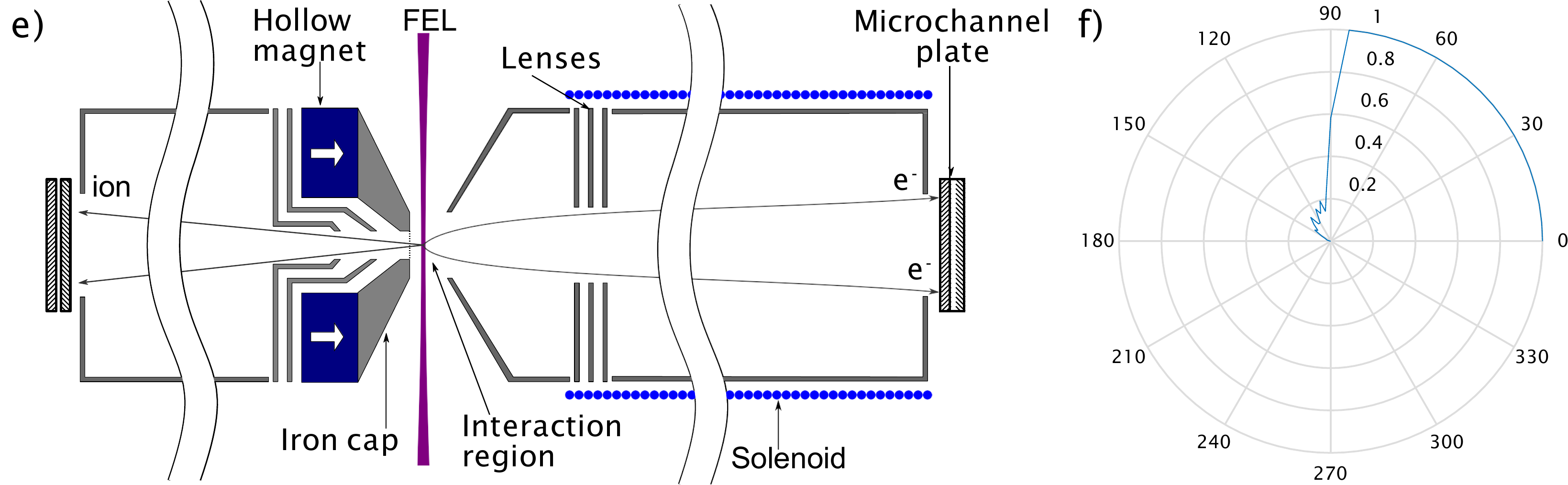}}
\caption{Free Electron Laser configuration for the generation of multiple harmonics and experimental setup. Configurations for the six undulators ($\mathrm{U_1-U_6}$)
for the generation of three (a) and four harmonics (b). In the first case, two undulators per harmonics were used, while in the second case, each harmonic was generated by
one undulator only. The phase-shifters ($\mathrm{PS_1-PS_5}$) used to control the relative phase between the harmonics are indicated in yellow for the two configurations.
Typical single-shot photoelectron spectra without (black lines) and with NIR pulse (red lines) measured for the three-(c) and four-(d) harmonic cases.
(e) Schematic, half-section view of the spectrometer, including the ion flight tube (left) and electron flight tube (right).
(f) Normalised simulated geometrical collection efficiency as a function of polar emission angle for 2-42 eV electrons for cylindrical magnet
configuration with the pole placed at 5 mm distance from interaction region. Electrons were simulated using steps of 5 eV. An emission angle of 0$^{\circ}$ (180$^{\circ}$)
corresponds to the axis of the spectrometer in (away from) the direction of the electron detector.
}
\label{Fig1n_ED}
\end{figure}

The experiment was performed at the seeded FEL FERMI and it is schematically presented in Fig.~\ref{Fig1n_ED}.
Two different configurations of the undulators were implemented for the generation of three (Fig.~\ref{Fig1n_ED} a) and four (Fig.~\ref{Fig1n_ED} b) harmonics.
The parameters of the FEL harmonics are reported in Tables~\ref{Table1n}.
\textcolor{black}{The seeding parameters (seed laser power and strength of the dispersive section) were carefully optimized in order
to produce a sufficiently high bunching
that could be preserved along the whole set of undulators tuned at the various harmonics. Tuning the first undulators to higher harmonics (shorter wavelengths)
and the later ones to the lower harmonics had a twofold motivation.
First, the bunching at higher harmonics was more prone to degradation and it would have been more difficult to preserve it up to the end of the undulator chain.
Second, the diffraction at longer wavelength was larger thus a shorter propagation path was preferable.
For a properly optimized setup each undulator group produced a coherent, $\sim$50-fs-long FEL pulse centered at the resonant wavelength.
The phase between the electric field of each harmonic was determined and controlled by the phase shifter available at FERMI at each undulator break.}

The XUV and NIR pulses (energy $E_{NIR}=45~\mathrm{\mu J}$, duration $\mathrm{FWHM}_{NIR}=60$~fs,
intensity $I_{NIR}=1.5\times10^{12}~\mathrm{W/cm^2}$) were temporally and spatially overlapped in the interaction region with a residual shot-to-shot delay jitter
of $\Delta\tau=\pm3$~fs using a recombination mirror for collinear propagation. In the interaction region a magnetic bottle electron spectrometer
(see Fig.~\ref{Fig1n_ED} e) collected the photoelectrons emitted in the upward hemisphere. The single-shot photoelectron spectra measured in neon with
(red lines) and without (black lines) the NIR pulse for the three and four harmonics configurations are shown in Fig.~\ref{Fig1n_ED} c, d, respectively.
For each machine setting, single-shot harmonic spectra (without NIR field) were measured~\cite{JSR-Zangrando-2015}.
From these data we estimated a typical shot-to-shot fluctuation (standard deviation) for the intensity of each harmonic of about $5\%-8\%$.
The energy of the single harmonic was proportional to the integral of the corresponding peak of the photoelectron spectrum.
The integral was corrected for the response function of the magnetic bottle spectrometer for different photoelectron energies,
the cross-section of the target gas, and the transmission of the XUV beamline~\cite{NATPHOT-Allaria-2012}.\\
The data were accumulated typically for $\approx10000-12000$ shots for each machine and phase setting.
For each setting, the mean intensities of the main photoelectron peaks $\langle I_q\rangle$ ($q=(7,8,9)$ or $q=(7,8,9,10)$) were determined.\\ 
The total energy of the FEL pulse was measured (on a single-shot basis) with an ionisation monitor placed upstream of the transmission and focusing XUV beamline.\\

\subsection*{Reconstruction of attosecond pulses using multi-IR photon transitions}
The temporal reconstruction of the attosecond pulse train using the correlation parameter $\rho_{q-1,q,q+1}$ for the simulations presented
in Fig.~\ref{Fig1}c-f is shown in Fig.~\ref{Fig8n_ED} a-d, respectively. The agreement between the input data (black curves) and the reconstructed profiles
(red (a), blue (b), green (c), and magenta(d)) indicates the validity of our reconstruction method based on the value of the correlation parameter $\rho_{7,8,9}$.
We also performed time-dependent Schr\"odinger Equation (TDSE) simulations which confirmed the validity of our reconstruction protocol.
The correlation curve obtained using the TDSE simulations reproduces that obtained by the strong field approximation (SFA) with a
small shift of 0.157~rad, which results in only minor corrections in the reconstruction of the intensity profile of the attosecond pulses.
This shift was taken into account in the reconstructions presented in this manuscript.

The validity of the temporal reconstruction of the attosecond pulse train from the shift of the oscillations of the sidebands
$S^{(+)}_{q-1,q}$ $\left(S^{(-)}_{q,q+1}\right)$ and $S^{(+)}_{q,q+1}$ $\left(S^{(-)}_{q,q+1}\right)$ is shown in Fig.~\ref{Fig8n_ED} e,
which reports the input (black lines) and reconstructed (blue dotted lines) intensity profiles for the simulation shown in Fig.~\ref{Fig1}b.

\subsection*{FEL simulations for attosecond pulse generation}
\textcolor{black}{The emission process for the configuration emitting the harmonics~7-8-9
can be simulated with the new version of the FEL code genesis~\cite{Genesis}. With the undulator set in the condition reported in
Fig.~\ref{Fig1n_ED}a the resonant wavelength to be followed by the FEL code is different in the three sets of undulators.
Given the fact that genesis only tracks a relatively narrow band field this requires that different simulations are performed for the different undulators sets.
This option is normally used for harmonic generation FEL schemes and has been largely used in the study of the HGHG operation mode implemented at FERMI.
The problem is here complicated by the fact that consecutive undulators are not tuned to wavelengths which are one harmonic of the other.
Thanks to the recent upgrade of genesis that allows tracking of each single electron in the beam this can be done if one carefully manages the transition
from one undulator set to the next. After the interaction with the external seed laser is simulated and the energy modulation at the 260~nm wavelength
is imprinted in the beam the particle phase-space is used to simulate the emission process at the 9th harmonic in the first group of undulators.
The electric field produced here is then propagated in free space to the exit of the whole radiator while particles are used for simulating the
emission process at the 8th harmonic in the second group of undulators. The same is done with field and particles at the exit of the second group of undulators.
Finally, three fields are produced representing the emission from each undulator set at the exit of the radiator. The result of the simulation is presented in Fig.~\ref{Fig2n_ED} a,b that
show the femtosecond envelope (a) and the attosecond structure (b) of the XUV pulse obtained using the combination of the three harmonics.
The resonance condition, phase shifter between sets, and other parameters can be adjusted as usual. Simulations rely on the standard electron beam and
seed laser for FERMI. Seed laser power and strength of the dispersive section are used as an optimization parameter to maximize the bunching and keep emission balanced
between various harmonics. The parameters used in the simulation are summarized in Table~\ref{Table4n}.}

\clearpage
\setcounter{page}{1}
\section*{Supplementary Information}
\subsection*{Supplementary Methods}
\subsubsection*{Magnetic bottle spectrometer}
The electron spectrometer is a magnetic bottle type electron spectrometer 
chosen for its unparalleled collection efficiency combined with good energy resolution over a broad range of electron kinetic energies. This enables
single-shot measurements with good photoelectron statistics. The operating principle is as follows:
by using a combination of a strong and weak magnetic fields, the electrons, which are initially emitted in all directions,
have their trajectories parallelized in the direction of decreasing magnetic field which ensures that almost all electrons are directed towards
the microchannel plate detector. This allows for a maximum collection solid angle of 4$\pi$~sr, and collection efficiencies of 50-60\%,
taking into account the detector efficiency. 
The particular spectrometer used for the experiment is an electron-ion coincidence spectrometer with a design similar
to that described previously.

To summarize, the strong magnetic field is created using a hollow cylindrical permanent magnet with peak field strength $\sim$1~T
in combination with a conical soft iron pole piece. The weak magnetic field is delivered using a solenoid of turn density~=~500~turns/m,
which extends from ~90 mm distance from the interaction region along the entire length of the $\sim$2~m long flight-tube,
terminating just before a 40~mm active diameter micro-channel plate detector (Hamamatsu F9892) placed at the exit of the flight tube.
A current of 0.5-2.0~A is sufficient to generate a magnetic field of a few~mT. At the entrance to the electron flight-tube,
a set of four electrostatic lenses with a circular aperture of 20~mm diameter and a spacing of 5 mm are used to apply a small potential across the
interaction region to ensure electrons close to zero kinetic energy arrive within a reasonable flight time.
Applying a potential to the final lens and the flight tube allows the electrons to be retarded by an arbitrary energy before entering the flight tube,
which gives improved resolution in the higher kinetic energy regions of the spectrum at the expense of detection capability for any electrons
with kinetic energy lower than the applied deceleration potential. Resolving powers ($E/\Delta E$) of 35 are possible.

Typically, a hollow cylinder magnet configuration is used so that coincident detection of photo-ions is also possible,
by accelerating them in the opposite direction of the electrons by applying a suitable potential to the electron flight tube entrance.
The ions fly through the hole in the magnet, are focused by a set of ion optics located behind the magnet, and then fly down a short flight
tube before being detected by an additional microchannel plate detector. A half-section schematic view of the entire setup including both electron and
ion flight tubes is shown in Fig.~\ref{Fig1n_ED} e. The ion detection configuration was used during optimization of the spectrometer and the
spatial overlap by looking at the charge distributions of multiple ionisation of rare gases.

In this instance however, the hollow magnet configuration's most critical property is that it reduces the collection efficiency to near
the 2$\pi$~sr required for this experiment, rather than the typical 4$\pi$ one can achieve with the setup.
This is accomplished by placing the magnet $\sim$5~mm away rather than the typical 10~mm away when in 4$\pi$ operation from the interaction volume. The result
of this change means that the source region is located at the peak of the magnetic field produced by the magnet, rather than being located on the gradient.
The magnetic mirror effect, which is responsible for reversing the electrons' direction in normal operation, relies on the electrons seeing an increasing
magnetic field. Hence, with the source region placed at the magnetic field maximum, electrons that are emitted towards the magnet will,
rather than be turned around and accelerated towards the electron detector, pass through the hole in the magnet.
They can then be prevented from reaching the electron detector by applying a moderate positive potential to the ion optical elements.
Electrons emitted away from the magnet are parallelized and detected normally. This behavior was confirmed using SIMION\textsuperscript{\textregistered}
simulations that showed that for $< 42$ eV electrons the detection efficiency for backwards propagating electrons is a few percent of that for
forward propagating electrons (see Fig.~\ref{Fig1n_ED} f).\\
\subsection*{Supplementary Equation(s)}
\subsubsection*{Strong field approximation model: multiple sideband generation}
The simulations of the photoelectron spectra generated by the combination of the XUV attosecond waveforms ($\mathbf{E}_{XUV}(t)$) and the NIR ($\mathbf{E}_{NIR}(t)$) pulses are based on the Strong Field Approximation (SFA).
The amplitude of the photoelectron wave packet emitted with final momentum $\mathbf{p}$ for a delay $\tau$ between the XUV and NIR pulses, $b(\mathbf{p},\tau)$, is given by (atomic units):
\begin{equation}
b(\mathbf{p},\tau)=i\int_{-\infty}^{+\infty} \mathbf{E}_{XUV}(t') \mathbf{d}\left[\mathbf{p}+\mathbf{A}(t')\right]\exp\left\{-i\int_{t'}^{+\infty}dt{''}\left[\frac{1}{2}(\mathbf{p}+\mathbf{A}(t^{''}))^2\right]+I_pt'\right\}dt'
\label{Eq-SFA}
\end{equation}
where $\mathbf{A}(t)$ is the vector potential of the NIR pulse, $t'$ the ionisation instant, $I_p$ the ionisation potential of the atom, $\mathbf{d}\left[\mathbf{p}+\mathbf{A}(t')\right]$ is the matrix element describing the transition from the ground state to the state with velocity $\mathbf{p}+\mathbf{A}(t')$ after the absorption of an XUV photon.\\
We simulated the photoelectron spectra generated in the two-color photoionisation using the following expression for the electric fields:
\begin{eqnarray} \label{eq:star1}
  \mathbf{E}_{XUV}(t) &=& \sum_{q}F_qf(t)\cos(q\omega_{UV}t+\varphi_q)\mathbf{u}_{XUV}\\
  \mathbf{E}_{NIR}(t-\tau) &=& F_{NIR}f_{NIR}(t-\tau)\sin(\omega_{NIR}(t-\tau))\mathbf{u}_{NIR}
  \label{eq:star2}
\end{eqnarray}
where $q$ is the harmonic order, $\omega_{UV}$ is the central frequency of the FEL seed laser, $F_q$ and $\varphi_q$ are the amplitude and phase of the $q$th harmonic, and $F_{NIR}$ and $\omega_{NIR}=\omega_{UV}/3$ are the amplitude and central frequency of the NIR pulse.
We assumed a gaussian temporal envelope common to all harmonics ($f(t)$) and for the NIR pulse ($f_{NIR}(t)$) with full-width at half maximum durations $\mathrm{FWHM}=50$~fs and $\mathrm{FWHM_{NIR}}=60$~fs, respectively.\\
 The polarisation directions of the two fields are indicated by the unit vectors $\mathbf{u}_{XUV}$ and $\mathbf{u}_{NIR}$. In the experiment, the fields were polarised along the same direction and parallel to the axis of the magnetic bottle electron spectrometer.

 Important information on the correlated variation of the sidebands can be derived analytically from Eq.~(\ref{Eq-SFA}). For the sake of simplicity, we neglect the slow intensity variation due to the envelopes $f(t)$ and $f_{NIR}(t)$. For low intensities of the NIR field, we can also neglect the term proportional to $A^{2}(t^{''})$ in Eq.~(\ref{Eq-SFA}). Moreover, we assume that the dipole matrix element is constant over the energy range spanned by the XUV spectrum ($\mathbf{d}[\mathbf{p}]=const.$) and we consider only electrons emitted parallel to the polarisation direction of the two fields: $\mathbf{p}\parallel \mathbf{u}_{XUV,NIR}\rightarrow\vartheta=0$.
These approximations allow one to identify the most relevant properties of the sidebands generated in the two-color field.
Using the following properties of the Bessel functions $J_n$:
\begin{eqnarray}\label{Eq-S3}
  e^{ix\sin\phi} &=& \sum_{n=-\infty}^{+\infty}J_n(x)e^{in\phi}\nonumber\\
  J_{-n}(x) &=&J_n(-x)=(-1)^nJ_{n}(x)
\end{eqnarray}
the amplitude of the photoelectron wave packets for two consecutive sidebands can be expressed as:
\begin{eqnarray}\label{Eq-S4}
  |b^{(-)}(\mathbf{p},\tau)|^2[\vartheta=0]\equiv S^{(-)}_{q,q+1}[\vartheta=0] &\propto&a^{(-)}_{q,q+1}-b^{(-)}_{q,q+1}\cos\left[\varphi_{q+1}-\varphi_{q}+3\omega_{NIR}\tau\right]\nonumber\\
  |b^{(+)}(\mathbf{p},\tau)|^2[\vartheta=0]\equiv S^{(+)}_{q,q+1}[\vartheta=0] &\propto&a^{(+)}_{q,q+1}+b^{(+)}_{q,q+1}\cos\left[\varphi_{q+1}-\varphi_{q}+3\omega_{NIR}\tau\right]
\end{eqnarray}
where:
\begin{eqnarray}\label{Eq-S5}
  a^{(-)}_{q,q+1}&\propto& J^2_1(x^{(-)}_{q,q+1})F_q^2+J^2_2(x^{(-)}_{q,q+1})F^2_{q+1}\nonumber\\
  a^{(+)}_{q,q+1}&\propto& J^2_2(x^{(+)}_{q,q+1})F_q^2+J^2_1(x^{(+)}_{q,q+1})F^2_{q+1}\nonumber\\
  b^{(-)}_{q,q+1}&\propto&2J_1(x^{(-)}_{q,q+1})J_2(x^{(-)}_{q,q+1})F_qF_{q+1}\nonumber\\
  b^{(+)}_{q,q+1}&\propto&2J_1(x^{(+)}_{q,q+1})J_2(x^{(+)}_{q,q+1})F_qF_{q+1}
\end{eqnarray}
\begin{eqnarray}
x^{(-)}_{q,q+1}&=& \sqrt{2[(q+1/3)\omega_{UV}-I_p]}F_{NIR}/\omega^2_{NIR}\nonumber\\
x^{(+)}_{q,q+1}&=& \sqrt{2[(q+2/3)\omega_{UV}-I_p]}F_{NIR}/\omega^2_{NIR}
\end{eqnarray}

Equation~(\ref{Eq-S4}) indicates that adjacent sidebands oscillate out-of-phase, independently from the relative phase of the harmonics (if $b^{(+)}_{q,q+1}b^{(-)}_{q,q+1}>0$) . 
In order to isolate the oscillating term in Eq.~(\ref{Eq-S4}), we define the sideband oscillation parameter as:
\begin{equation}\label{Eq-S6}
  P_{q,q+1}=\frac{S^{(+)}_{q,q+1} -S^{(-)}_{q,q+1}}{S^{(+)}_{q,q+1} +S^{(-)}_{q,q+1}}
\end{equation}
The definition of the parameter $P_{q,q+1}$ is equivalent to the asymmetry parameter discussed in several photoionisation experiments. 
In the case of equal amplitudes of the two harmonics ($F_q=F_{q+1}$) the expression for the parameter becomes:
\begin{equation}\label{Eq-SP7}
 P_{q,q+1}=\frac{\left(a^{(+)}_{q,q+1}-a^{(-)}_{q,q+1}\right)+\left(b^{(+)}_{q,q+1}+b^{(-)}_{q,q+1}\right)\cos\left[\varphi_{q+1}-\varphi_{q}+3\omega_{NIR}\tau\right]}{\left(a^{(+)}_{q,q+1}+a^{(-)}_{q,q+1}\right)+\left(b^{(+)}_{q,q+1}-b^{(-)}_{q,q+1}\right)\cos\left[\varphi_{q+1}-\varphi_{q}+3\omega_{NIR}\tau\right]}
\end{equation}
In our experimental conditions (for the intensity of the NIR pulse and for the momentum $|\mathbf{p}|$ of the electron) and using the explicit expression of the Bessel functions, we can make the approximations:
\begin{eqnarray*}\label{Eq-S7b}
  a^{(+)}_{q,q+1} &\approx& a^{(-)}_{q,q+1} \\
  b^{(+)}_{q,q+1} &\approx& b^{(-)}_{q,q+1}
\end{eqnarray*}
and Eq.~(\ref{Eq-SP7}) can be written as:
\begin{equation}\label{Eq-S8}
 P_{q,q+1}=\alpha_{q,q+1}\cos\left[\varphi_{q+1}-\varphi_{q}+3\omega_{NIR}\tau\right]
\end{equation}
where $\alpha_{q,q+1}=b^{(+)}_{q,q+1}/a^{(+)}_{q,q+1}$.\\
The correlation plots of the $P_{q,q+1}$ and $P_{q-1,q}$ parameters are ellipses described by the equation:
\begin{equation}\label{Eq-S9}
 \frac{P^2_{q-1,q}}{\alpha^2_{q-1,q}}+ \frac{P^2_{q,q+1}}{\alpha^2_{q,q+1}}-2\frac{P_{q-1,q}}{\alpha_{q-1,q}}\frac{P_{q,q+1}}{\alpha_{q,q+1}}\cos\Delta\varphi_{q-1,q,q+1}=\sin^2\Delta\varphi_{q-1,q,q+1}
\end{equation}
where:
\begin{equation}\label{Eq-S10}
\Delta\varphi_{q-1,q,q+1}=\left(\varphi_{q+1}-\varphi_{q}\right)-\left(\varphi_{q}-\varphi_{q-1}\right)=\varphi_{q+1}+\varphi_{q-1}-2\varphi_{q}
\end{equation}
We can further simplify this expression by using the approximation $\alpha=\alpha_{q,q+1}\approx\alpha_{q-1,q}$ and we obtain:
\begin{equation}\label{Eq-S11}
P^2_{q-1,q}+P^2_{q,q+1}-2P_{q-1,q}P_{q,q+1}\cos\Delta\varphi_{q-1,q,q+1}=\alpha^2\sin^2\Delta\varphi_{q-1,q,q+1}
\end{equation}
 Equation~(\ref{Eq-S11}) represents an ellipse, whose major axis is oriented at an angle $\beta=+45^{\circ}$ $(-45^{\circ})$ for values $0\leq\Delta\varphi_{q-1,q,q+1}<\pi/2;3\pi/2\leq\Delta\varphi_{q-1,q,q+1}<2\pi $ ( $\pi/2\leq\Delta\varphi_{q-1,q,q+1}<3\pi/2$).

\begin{figure}[hb]
\centering
\includegraphics[width=1.0\linewidth, height=10cm]{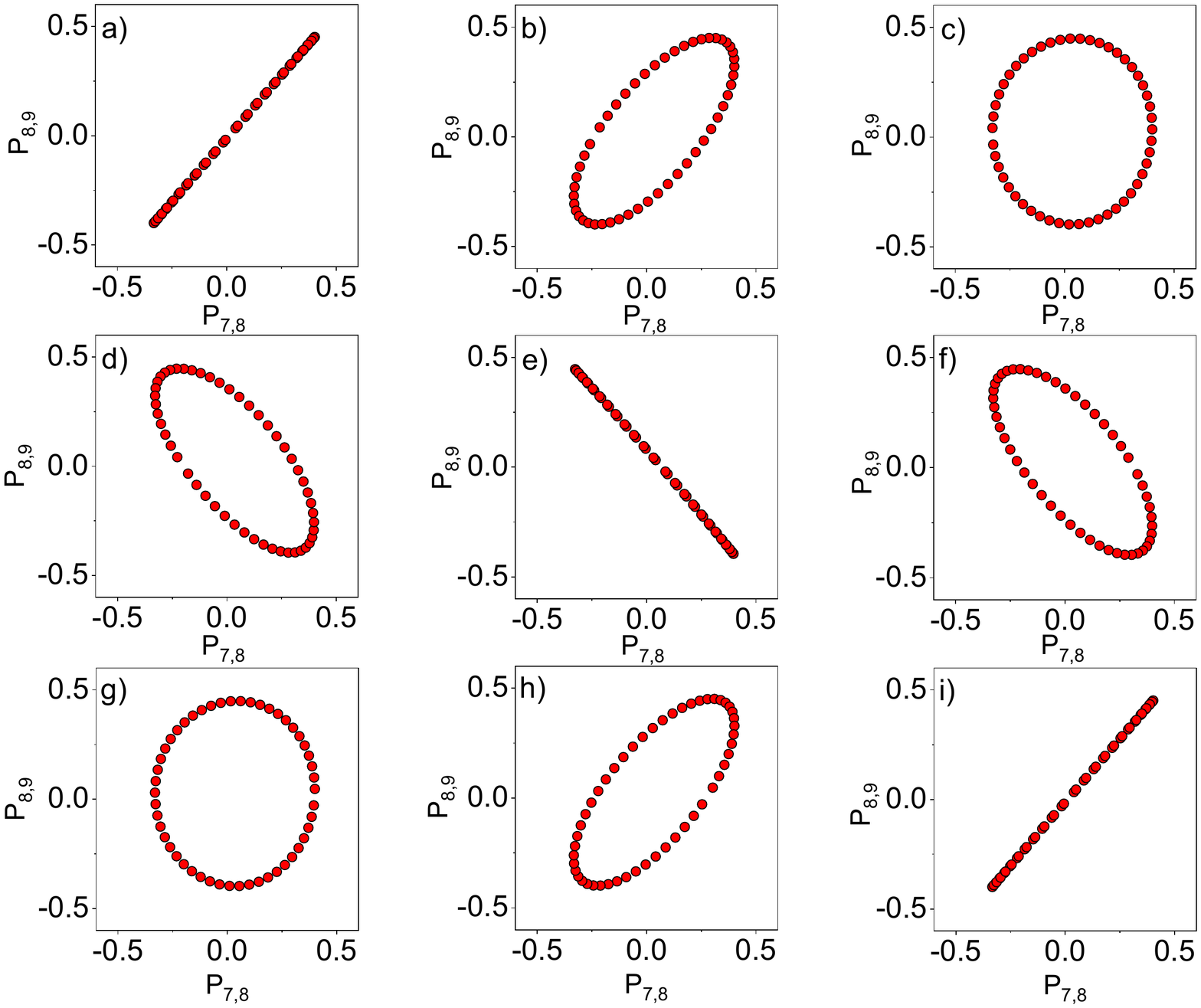}
\caption{Simulated correlation plots $\left(P_{8,9},P_{7,8}\right)$ for different values of $\Delta\varphi_{7,8,9}$ from $0$ to $2\pi$ in steps of $\pi/4$: $\Delta\varphi_{7,8,9}=0$ (a), $\pi/4$ (b), $\pi/2$ (c), $3\pi/4$ (d), $\pi$ (e), $5\pi/4$ (f), $3\pi/2$ (g), $7\pi/4$ (h), and , $2\pi$ (i) . The intensities of the three harmonics are equal.}
\label{Fig4n_ED}
\end{figure}

Figure~\ref{Fig4n_ED} shows the simulated correlation plots $\left[P_{q-1,q}, P_{q,q+1}\right]$ for nine different values of the $\Delta\varphi_{q-1,q,q+1}$ in steps of $\pi/4$.
As the difference of the relative phases $\Delta\varphi_{q-1,q,q+1}$ changes from $0$ to $\pi/2$ the major axis is oriented at $\beta=45^{\circ}$ and the pattern changes from a line ($\Delta\varphi_{q-1,q,q+1}=0$) (a), to an ellipse ($0<\Delta\varphi_{q-1,q,q+1}<\pi/2$) (b) and to a circle ($\Delta\varphi_{q-1,q,q+1}=\pi/2$) (c). For $\pi/2<\Delta\varphi_{q-1,q,q+1}<3\pi/2$ the major axis of the pattern is oriented at $\beta=-45^{\circ}$ and we have a transition from an ellipse ($\pi/2<\Delta\varphi_{q-1,q,q+1}<\pi$) (d), to a line oriented at $-45^{\circ}$ (e), back to an ellipse ($\pi<\Delta\varphi_{q-1,q,q+1}<3\pi/2$) (f), and finally to a circle ($\Delta\varphi_{q-1,q,q+1}=3\pi/2$) (g). For larger phase differences ($3\pi/2<\Delta\varphi_{q-1,q,q+1}<2\pi$) the ellipse major axis is oriented at $\beta=+45^{\circ}$ (h) and it reduces back to a line ($\Delta\varphi_{q-1,q,q+1}=2\pi$) (i).

\textcolor{black}{The distributions presented in Fig.~\ref{Fig4n_ED} where obtained using Eq.~(\ref{Eq-S8}) and a constant delay difference between consecutive points of $\Delta\tau=17.7$ as in the interval $[0; 2\pi/(3\omega_{NIR})]$.
As it can be observed the density of points increases in correspondence of the major axis of the ellipses (i.e. for those regions that present the largest distance from the origin of the plane $(P_{7,8},P_{8,9})$). This geometrical
property of the distributions can be recognized in the experimental data presented in the main manuscript in Figs.~\ref{Fig2},\ref{Fig3},\ref{Fig4}. In the experiment the delay between the attosecond waveform and NIR pulses follows a gaussian distribution (FWHM= 6 fs)}.

From Eq.~(\ref{Eq-SFA}) and Eq.~(\ref{Eq-S3}), it is straightforward to demonstrate that sidebands emitted in the opposite directions $\vartheta=0$ (see Eq.~(\ref{Eq-S4})) and $\vartheta=\pi$ along the polarisation direction of the field oscillates with a $\pi$ phase offset:
\begin{eqnarray}\label{Eq-S12}
  |b^{(-)}(\mathbf{p},\tau)|^2[\vartheta=\pi]\equiv S^{(-)}_{q,q+1}[\vartheta=\pi] &\propto&a^{(-)}_{q,q+1}+b^{(-)}_{q,q+1}\cos\left[\varphi_{q+1}-\varphi_{q}+3\omega_{NIR}\tau\right]\nonumber\\
  |b^{(+)}(\mathbf{p},\tau)|^2[\vartheta=\pi]\equiv S^{(+)}_{q,q+1}[\vartheta=\pi] &\propto&a^{(+)}_{q,q+1}-b^{(+)}_{q,q+1}\cos\left[\varphi_{q+1}-\varphi_{q}+3\omega_{NIR}\tau\right]
\end{eqnarray}
As a consequence integration of the entire volume ($4\pi$~sr) of the photoelectron spectrum cancels out the sideband oscillations as a function of the relative delay $\tau$. This is a direct consequence of the opposite symmetry of the final continuum state reached by the two different pathways leading to the same final energy:
$(A^{(+1)}_q$ and $A^{(-2)}_{q+1};A^{(+2)}_q$ and $A^{(-1)}_{q+1})$ (see Eq.~(1) of the main manuscript). The opposite symmetry is the result of the different number of photons involved in the two pathways.\\
For the experiment the collection angle of the magnetic bottle was fixed to $\approx2\pi$~sr along the polarisation direction of the two fields to avoid the cancellation of the oscillations in the sidebands photoelectron peaks.
\\


\subsubsection*{Perturbation theory approach}\label{sec:pt}
The present version of the SFA describes the electron in the continuum as a plane wave,
neglecting its interaction with the residual ion.
It is instructive to consider possible effects of this interaction.
We do so here within the lowest nonvanishing order of perturbation theory (PT)
for the pulses containing many optical cycles and switching on and off adiabatically.
Taking into account two pathways and using standard methods, one can obtain
the photoelectron angular distributions of the sidebands in the form of a partial wave expansion
and the Legendre polynomials $P_{\kappa}(x)$:
\begin{eqnarray} \label{eq:padminus}
S_{q,q+1}^{(-)}(\vartheta)  & = & \sum_{\kappa}
\sum_{\ell L \atop \ell' L'} Z_{\kappa}(\ell \ell' L L') \,
 \bigg( A_q^{(+1)}(\ell L) A_q^{(+1) \, \ast}(\ell' L') \nonumber \\
 & & +  A_{q+1}^{(-2)}(\ell L) A_{q+1}^{(-2) \, \ast}(\ell' L') +
    2 \Re \left[ A_q^{(+1)} (\ell L) \, A_{q+1}^{(-2) \, \ast}(\ell' L') \right] \bigg)
  P_{\kappa}(\cos \vartheta) \,,
\end{eqnarray}
\begin{eqnarray} \label{eq:padplus}
S_{q,q+1}^{(+)}(\vartheta) & = &  \sum_{\kappa}
\sum_{\ell L \atop \ell' L'} Z_{\kappa}(\ell \ell' L L') \,
 \bigg( A_q^{(+2)}(\ell L) A_q^{(+2) \, \ast}(\ell' L') \nonumber \\
 & & +  A_{q+1}^{(-1)}(\ell L) A_{q+1}^{(-1) \, \ast}(\ell' L') +
    2 \Re \left[ A_q^{(+1)} (\ell L) \, A_{q+1}^{(-2) \, \ast}(\ell' L') \right] \bigg)
  P_{\kappa}(\cos \vartheta) \,.
\end{eqnarray}
Here $A_q^{(\pm n)} (\ell L)$ denotes amplitude of ionisation
from the ground atomic state to a channel
with the photoelectron orbital momentum $\ell$ and total orbital
momentum of the system 'photoelectron + ion' $L$ ($\bf{L} = \bf{L}_f + \bf{\ell}$,
with $L_f = 1$ being the orbital momentum of the residual ion)
by the XUV harmonic $q$ and either absorption $(+n)$, or emission $(-n)$ of $n$ photons
($n=1, \,2$). The amplitudes with $n=1$ or $n=2$ are the second or the third order
amplitudes, respectively. We introduce real numerical coefficients
\begin{eqnarray} \label{eq:z}
Z_{\kappa}(\ell \ell' L L') &  = &  Z_{\kappa}(\ell' \ell L' L) \hspace{45mm} \\
  & = & \minus{L_f} \hat{\ell} \hat{\ell}' \hat{L} \hat{L}' \CGC{\ell}{0}{\ell'}{0}{\kappa}{0} \,
\CGC{L}{0}{L'}{0}{\kappa}{0} \SechsJ{\ell}{\ell'}{\kappa}{L'}{L}{L_f} \,,  \nonumber
\end{eqnarray}
where $\hat{a} = \sqrt{2a+1}$ and standard notation is used for the Clebsch-Gordan
coefficients and Wigner 6$j$-symbol. After integrating~(\ref{eq:padminus}) and
(\ref{eq:padplus}) over a hemisphere $0 \leq \vartheta \leq \pi/2$
only terms with odd and zero $\kappa$ are left.
Terms with odd $\kappa$ contain only interference between second and
third order amplitudes.
Furthermore, we factor out the phases of the fields from the amplitudes. For the fields defined
by Eqs.~(\ref{eq:star1}), (\ref{eq:star2}):
\begin{equation} \label{eq:u2plus}
A_q^{(\pm n)}(\ell L) =  \frac{F_q}{2} \left( \frac{F_{NIR}}{2} \right)^n  e^{i(\varphi_q \mp n \varphi_{NIR})} \,
c_q^{(\pm n)}(\ell L) \,,
\end{equation}
where $c_q^{(\pm n)}(\ell L)$ are independent of the field parameters and should be generally
calculated within an atomic model. As a result,
the intensities of the sidebands (\ref{eq:padminus}), (\ref{eq:padplus})
can be cast into the form (compare with Eq.~(1) of the main manuscript)
\begin{eqnarray} \label{eq:elli1}
S_{q,q+1}^{(-)} & = &  \sigma_{q,q+1}^{(-)} + \left| X_{q,q+1}^{(-)} \right|
 \cos (\varphi_{q+1} - \varphi_q + 3 \varphi_{NIR} - \xi_{q,q+1}^{(-)}  ) , \\
S_{q,q+1}^{(+)} & = &  \sigma_{q,q+1}^{(+)} + {\left| X_{q,q+1}^{(+)} \right|}
\cos (\varphi_{q+1} - \varphi_q + 3 \varphi_{NIR} + \xi_{q,q+1}^{(+)} ) , \label{eq:elli2}
\end{eqnarray}
where $\varphi_{NIR} = \omega_{NIR} \tau$,
\begin{eqnarray} \label{eq:s1}
\sigma_{q,q+1}^{(-)} &  = & \sum_{\ell L} \left| A_q^{(+1)}(\ell L) \right|^2 +
\sum_{\ell L} \left|  A_{q+1}^{(-2)}(\ell L)    \right|^2 \,, \\
\sigma_{q,q+1}^{(+)} &  = & \sum_{\ell L} \left| A_q^{(+2)}(\ell L) \right|^2 +
\sum_{\ell L} \left|  A_{q+1}^{(-1)}(\ell L)    \right|^2  \label{eq:s2}
\end{eqnarray}
is a ``half cross section'' of excitation of the sidebands,
\begin{eqnarray} \label{eq:xa}
X_{q,q+1}^{(-)} &  = & \sum_{\ell'>\ell \atop L' \neq L} B(\ell \ell' L L') \,
c_q^{(+1)}(\ell L) \, c_{q+1}^{(-2) \, \ast}(\ell' L') , \\
X_{q,q+1}^{(+)} &  = & \sum_{\ell'>\ell \atop L' \neq L} B(\ell \ell' L L') \,
c_{q+1}^{(-1)}(\ell L) \, c_q^{(+2) \, \ast}(\ell' L') . \label{eq:xb} \,,
\end{eqnarray}
\begin{equation} \label{eq:bk}
B(\ell \ell' L L') = \frac{1}{16} F_{NIR}^3 F_q F_{q+1} \sum_{\kappa=1,3,5} C_{\kappa}
Z_{\kappa}(\ell \ell' L L') \, ,
\end{equation}
\begin{equation} \label{eq:cossin}
\cos \, \xi_{q,q+1}^{(\pm)}  =  \frac{ \Re \, X_{q,q+1}^{(\pm)} }{\left| X_{q,q+1}^{(\pm)} \right|}, \quad
\sin \, \xi_{q,q+1}^{(\pm)}  =  \frac{ \Im \, X_{q,q+1}^{(\pm)} }{\left| X_{q,q+1}^{(\pm)} \right|} \,,
\end{equation}
$C_1 = \frac{1}{2}$, $C_3 = - \frac{1}{8}$, $C_5 = \frac{1}{16}$.
The phases $\xi^{(\pm)}_{q,q+1}$ are due to the photoelectron-ion interaction and are usually indicated as atomic phases.
To proceed further we use relationships between amplitudes of photon absorption and emission.
with the additional assumption of the flat continuum
and assume equal intensities of the harmonics. Then
\begin{eqnarray} \label{eq:xx}
\left| X_{q,q+1}^{(+)} \right| & = & \left| X_{q,q+1}^{(-)} \right| =
\left| X_{q-1,q}^{(+)} \right| = \left| X_{q-1,q}^{(-)} \right|  \equiv X > 0 \,, \\
\sigma^{(+)}_{q,q+1} & = & \sigma^{(-)}_{q,q+1} = \sigma^{(+)}_{q-1,q} =
\sigma^{(-)}_{q-1,q} \equiv \sigma > 0 \,, \label{eq:ss} \\
\xi^{(+)}_{q,q+1} & = & \xi^{(+)}_{q-1,q} \equiv \xi, \qquad
\xi^{(-)}_{q,q+1} = \xi^{(-)}_{q-1,q} = \pi - \xi \,.
\label{eq:xixi}
\end{eqnarray}
With Eqs.~(\ref{eq:xx}-\ref{eq:xixi}) formulas (\ref{eq:elli1}), (\ref{eq:elli2}) turn into Eq.~(1)
of the SFA, provided the atomic phase is zero.

The flat-continuum approximation in our case means a negligible difference between
the relative values of the partial transition amplitudes
$c_{q-1}^{(+n)}$, $c_{q}^{(+n)}$, and $c_{q+1}^{(+n)}$ and similarly between $c_{q-1}^{(-n)}$, $c_{q}^{(-n)}$, and $c_{q+1}^{(-n)}$.
This is a good approximation for atomic neon, which does not contain features such
as resonances and Cooper minima in the spectral interval of our interest.
In argon, for example, close to the ionisation threshold the relative amplitudes
are much more sensitive to the energy and need to be calculated when treating the
high harmonic generation.

For the sideband oscillation parameter~(\ref{Eq-S6}) we obtain after transformations
\begin{equation} \label{eq:pqq}
P_{q,q+1} = \sigma^{-1} X \, \cos (\varphi_{q+1} - \varphi_q - 3 \varphi_{NIR} + \xi ) \,,
\end{equation}
where $ | \sigma^{-1} X | < 1 $. Thus we obtain the shape of the
correlation plot [$P_{q-1,q}, \, P_{q,q+1}$] identical to that obtained in the
SFA (see Eq.~(\ref{Eq-S8})).
Note that neither the sign of $\sigma^{-1} X$, nor the additional
atomic phase $\xi$, influence the plot, because both can be included in
a rescaled phase $\varphi_{NIR}$. Furthermore, Eqs.~(\ref{eq:elli1}), (\ref{eq:elli2}) remain
valid without integrating over the hemisphere, provided $\sigma^{(-)}_{q,q+1}$,
$\sigma^{(+)}_{q,q+1}$, and coefficients $C_{\kappa}$ in Eq.~(\ref{eq:bk}) are
changed accordingly. Therefore the correlation plot [$P_{q-1,q}, \, P_{q,q+1}$] remains
the same for arbitrary electron emission angle $\vartheta$ (including $\vartheta=0$, as in
the derivation in the SFA).\\

\subsubsection*{Time-dependent Schr\"odinger equation}\label{TDSE}
\begin{figure}[hb]
\centering
\includegraphics[width=0.9\textwidth]{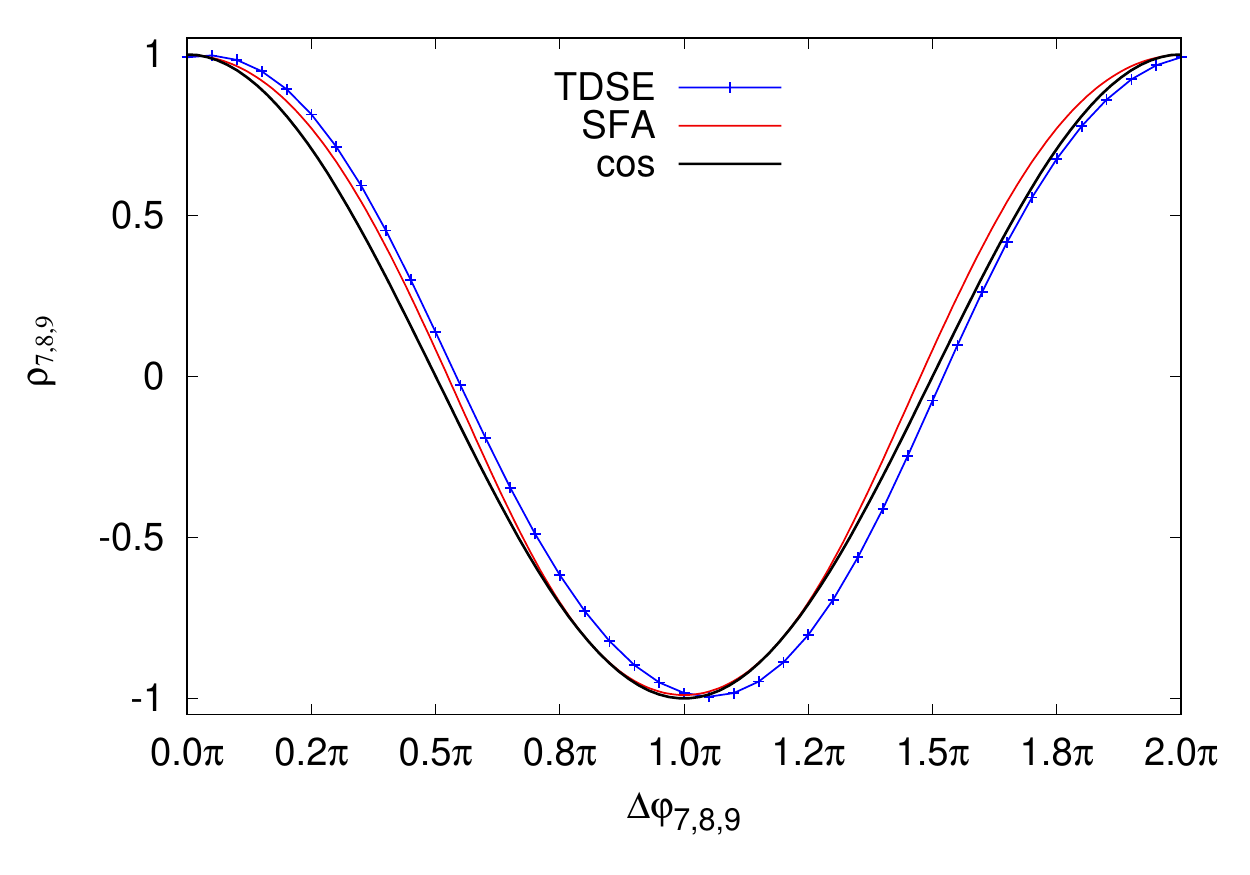}
\caption{Evolution of the correlation parameter $\rho_{7,8,9}$ as a function of the phase difference $\Delta\varphi_{7,8,9}$ simulated using the SFA approximation (red) and the TDSE (blue). The black curve indicates a cosine evolution.}
\label{Fig5n_ED}
\end{figure}

Numerical solutions of time-dependent Schr\"odinger equation (TDSE) were calculated using the single active electron model for neon atoms subjected to a combination of XUV and NIR pulses.
In the calculations the NIR pulse envelope is a trapezoid with six cycles ($6T_{NIR}$) of constant intensity ($I_{NIR}=2\times10^{12}~\mathrm{W/cm^2}$) and one cycle ramps. The XUV pulse has sine-squared envelope and a full duration of 5~NIR cycles (FWHM of approximately 5~fs).
The attosecond pulse train consists of harmonics 7-10 of the seed wavelength, with a phase added to harmonic 9. The final state wave function was analyzed as a function of energy and XUV-NIR delay using a standard window-operator method 
and the sideband populations were summed to construct sideband functions $P_{q,q+1}$ (Eq.~3 of main manuscript). We then constructed the correlation parameter $\rho_{7,8,9}$ as a function of the phase added to the 9th harmonic using Eq.~\ref{Eq-S13}. The result is shown in Fig.~\ref{Fig5n_ED}, which reports the correlation parameter $\rho_{7,8,9}$ determined using the SFA approximation (red) and the TDSE (blue). Both curves follow an evolution close to a cosine function, with an offset of about 0.157~rad. The offset is due to the different atomic phase accumulated by the two pathways leading to the same sideband. The effect of the shift was considered in the reconstruction of the intensity profile of the attosecond pulses.

\subsection*{Supplementary Notes}

\subsubsection*{Phase Scan: correlation plots and correlation coefficient}
For a fixed setting of the phase shifters, we indicate with $S^{(\pm)}_{q,q+1}(i)$ ($i=1,2,3,...n$) the intensity of the sideband $S^{(\pm)}_{q,q+1}$ for the $i$th-FEL shot.
The intensity of each sideband was obtained by numerical integration of the corresponding peak in the photoelectron spectrum. The sideband oscillation parameter $P_{q,q+1}$ was determined according to Eq.~(\ref{Eq-S6}).\\
The correlation coefficient $\rho_{q-1,q,q+1}$ is defined as:
\begin{equation}\label{Eq-S13}
\rho_{q-1,q,q+1}=\frac{\mathrm{cov}(x,y)}{\sigma_x\sigma_y}=\frac{\langle xy\rangle-\langle x\rangle\langle y\rangle}{\sqrt{\langle x^2\rangle-\langle x\rangle^2}\sqrt{\langle y^2\rangle-\langle y\rangle^2}}\quad x\equiv P_{q-1,q}(i)\quad y\equiv P_{q,q+1}(i)
\end{equation}
where:
\begin{equation}\label{Eq-S14}
\langle x\rangle=\frac{\sum_{i=1}^{N}x(i)}{N}\quad i=1,2,3,...n
\end{equation}
A similar definition was used for $\langle y\rangle$.
Perfect positive (negative) correlation $\rho_{q-1,q,q+1}=+1$ $(-1)$ indicates that $\Delta\varphi_{q-1,q,q+1}=0$ $(\pi)$.
Correlation $\rho_{q-1,q,q+1}=0$ indicates that $\Delta\varphi_{q-1,q,q+1}=\pi/2$ or $\Delta\varphi_{q-1,q,q+1}=3\pi/2$.
In general, each value of the correlation coefficient corresponds to two different phases $\Delta\varphi_{q-1,q,q+1}=\varphi_0$ and $\Delta\varphi_{q-1,q,q+1}=2\pi-\varphi_0$ (or $\Delta\varphi_{q-1,q,q+1}=-\varphi_0$).
This ambiguity implies that the shape of the ellipse (see Eq.~(\ref{Eq-S11})) does not permit distinction over the direction of the time $t$ i.e,
it does not allow us to distinguish between $E(t)$ and $E(-t)$.
This ambiguity can be resolved by observing that, starting from the conditions $\Delta\varphi_{q-1,q,q+1}=0$ (harmonics are in phase),
corresponding to the maximum positive correlation (see Fig.~\ref{Fig4n_ED}, a and i), the sign (positive or negative) of the delay introduced in the
propagation of the $q$th harmonic by the phase shifter is known. By measuring a complete phase-scan,
it is then possible to determine the sign of the phase difference $\Delta\varphi_{q-1,q,q+1}$ and to reconstruct without any ambiguity the attosecond waveform.
We have verified that the correspondence between phase difference and correlation parameter depends only slightly on the intensity of the NIR pulse
(in the perturbative regime) and on the momentum $\mathbf{p}$ of the electron.

The range of oscillation of the correlation coefficient $\rho_{7,8,9}$ is less than two for the measurement presented in Fig.~3 of the main manuscript. We attribute this reduced contrast to the single-shot intensity fluctuations of a single harmonic, which can introduce an additional (positive or negative) correlation between the shot-to-shot variations of the sideband intensities. In the XUV-NIR experiment we characterised the single-shot total intensity fluctuation, but we had no access to the single-shot intensity measurement of each individual harmonic.\\
Other possible sources for the reduction of the interval of the correlation coefficient are the averaging over the interaction volume and integration of the photoelectron spectrum over the solid angle of $2\pi$. All these effects will be considered in more detail in a subsequent manuscript.

In the experiment, the correlation coefficient evolution as a function of the phase difference $\Delta\varphi_{q-1,q,q+1}$ was fitted by a sinusoidal curve as shown in Fig.~3 of the main manuscript:
\begin{equation}\label{Eq-S15}
\rho_{q-1,q,q+1}=B_{q-1,q,q+1}\cos(\Delta\varphi_{q-1,q,q+1})+C_{q-1,q,q+1}
\end{equation}
where $B_{q-1,q,q+1}$ indicates the amplitude of the oscillation of the correlation coefficient and the offset $C_{q-1,q,q+1}$ takes into account possible additional correlations (negative or positive) between the fluctuations of the sideband intensities.\\
The error bar on the correlation coefficient was estimated as the standard deviation ($\sigma(\rho_{q-1,q,q+1})$) over several measurements (typically ten) acquired for the same experimental settings (amplitude and phases of the FEL harmonics and parameters of the NIR field).

\subsubsection*{Determination of the phase difference $\Delta\varphi_{q-1,q,q+1}$}
We describe the four harmonics used in the experiment with the following expressions:
\begin{eqnarray}\label{Eq-S16}
  E_{10}(t) &=&\frac{1}{2}\left\{\tilde{E}_{10}(t)+\tilde{E}^*_{10}(t)\right\}=\frac{1}{2}\left\{F_{10}f(t)\exp\left\{i\left[10\omega_{UV}t+\varphi^{(0)}_{10}\right]\right\}+c.c\right\}\nonumber \\
  E_{9}(t) &=&\frac{1}{2}\left\{\tilde{E}_{9}(t)+\tilde{E}^*_{9}(t)\right\}=\frac{1}{2}\left\{F_{9}f(t)\exp\left\{i\left[9\omega_{UV}t+\varphi^{(0)}_9\right]\right\}+c.c.\right\}\nonumber\\
  E_{8}(t) &=&\frac{1}{2}\left\{\tilde{E}_{8}(t)+\tilde{E}^*_{8}(t)\right\}=\frac{1}{2}\left\{F_{8}f(t-\tau_{sx})\exp\left\{i\left[8\omega_{UV}(t-\tau_{sx})+\varphi^{(0)}_8\right]\right\}+c.c.\right\}\nonumber\\
  E_{7}(t) &=&\frac{1}{2}\left\{\tilde{E}_{7}(t)+\tilde{E}^*_{7}(t)\right\}=\frac{1}{2}\left\{F_{7}f(t-\tau_{sx}-\tau_{sy})\exp\left\{i\left[7\omega_{UV}(t-\tau_{sx}-\tau_{sy})+\varphi^{(0)}_7\right]\right\}+c.c.\right\}\nonumber\\
\end{eqnarray}
where $\tau_{sx}$ and $\tau_{sy}$ are the time delays introduced by the phase shifters $\mathrm{PS_2}$ ($x=2$) and $\mathrm{PS_4}$ ($y=4$), and $\mathrm{PS_2}$ ($x=2$) and $\mathrm{PS_3}$ ($y=3$) for the three and four-harmonic case, respectively (see Fig.~2 of the main manuscript). For the three-harmonic case $F_{10}=0$. $\varphi^{(0)}_{q}$ indicates the initial phase of the $q$th harmonic, which is related to the phases $\varphi_q$ by the relations:
\begin{equation}
  \varphi_{10}=\varphi_{10}^{(0)};\quad\varphi_{9}=\varphi_{9}^{(0)};\quad\varphi_{8}=\varphi_{8}^{(0)}-8\omega_{UV}\tau_{sx};\quad\quad\varphi_{7}=\varphi_{7}^{(0)}-7\omega_{UV}(\tau_{sx}+\tau_{sy})
\end{equation}
It is important to observe that, due to the geometry of the undulators setup, the delay $\tau_{sx}$ introduced by the phase shifter $\mathrm{PS_2}$ affects the phases of both the 8th and 7th harmonic, while the delay $\tau_{sy}$ introduced by the phase shifter $\mathrm{PS_3}$ or $\mathrm{PS_4}$ affects the phase of the 7th harmonic only.
From Eq.~(\ref{Eq-S16}) the phase difference between the two paths contributing to the sideband $S^{(-)}_{q,q+1}$ is given by:
\begin{eqnarray*}\label{Eq-S17}
    \delta\varphi_{10-9}&=& \left(\varphi^{(0)}_{10}-\varphi^{(0)}_9\right)+3\omega_{NIR}\tau\\
    \delta\varphi_{9-8} &=& \left(\varphi^{(0)}_9-\varphi^{(0)}_8\right)+3\omega_{NIR}\tau+8\omega_{UV}\tau_{sx} \\
    \delta\varphi_{8-7} &=& \left(\varphi^{(0)}_8-\varphi^{(0)}_7\right)+3\omega_{NIR}(\tau-\tau_{sx})+7\omega_{UV}\tau_{sy}
\end{eqnarray*}
The phase differences $\Delta\varphi_{7,8,9}$ and $\Delta\varphi_{8,9,10}$ are expressed by the relations:
\begin{eqnarray}\label{Eq-calibration}
  \Delta\varphi_{8,9,10} &=& \delta\varphi_{10-9}-\delta\varphi_{9-8}= \left(\varphi^{(0)}_{10}+\varphi^{(0)}_{8}-2\varphi^{(0)}_9\right)-8\omega_{UV}\tau_{sx}\nonumber\\
  \Delta\varphi_{7,8,9} &=& \delta\varphi_{9-8}-\delta\varphi_{8-7}= \left(\varphi^{(0)}_{9}+\varphi^{(0)}_{7}-2\varphi^{(0)}_8\right)+9\omega_{UV}\tau_{sx}-7\omega_{UV}\tau_{sy}
\end{eqnarray}
Therefore the phase difference $\Delta\varphi_{8,9,10}$ will change as a function of the delay $\tau_{sx}$ only due to the delay introduced on the 8th harmonic.
In the case of the phase difference $\Delta\varphi_{7,8,9}$ we have two different terms related to $\tau_{sx}$ and $\tau_{sy}$. The first one takes into account that a delay $\tau_{sx}$ will delay both the 8th and 7th harmonic, which is equivalent to a shift in time of the 9th harmonic in the opposite direction ($-\tau_{sx}$) . The second term describes the additional phase accumulated by the 7th harmonic due to the delay $\tau_{sy}$.\\
Using a calibration curve in the $\tau_{sx}$ and one in $\tau_{sy}$ (for a generic, fixed value of $\tau_{sx}$), the phase differences for a generic combination of $\tau_{sx}$ and $\tau_{sy}$ can be determined using Eq.~(\ref{Eq-calibration}).

The error interval for the phase difference $\Delta\varphi_{q-1,q,q+1}$ was estimated as:
\begin{equation}
\sigma(\Delta\varphi_{q-1,q,q+1})=\frac{\sigma(\rho_{q-1,q,q+1})}{|B_{q-1,q,q+1}\sin\Delta\varphi_{q-1,q,q+1}|}
\end{equation}
Around $\Delta\varphi_{q-1,q,q+1}\approx0,\pi,2\pi$ this formula cannot be applied and we used the next term in the Taylor expansion of Eq.~(\ref{Eq-S15}), which leads to the error:
\begin{equation}
\sigma(\Delta\varphi_{q-1,q,q+1})=\sqrt{\frac{2\sigma(\rho_{q-1,q,q+1})}{|B_{q-1,q,q+1}\cos\Delta\varphi_{q-1,q,q+1}|}}
\end{equation}

The correlation coefficients $\rho_{7,8,9}$ and phase differences $\Delta\varphi_{7,8,9}$ for the measurements shown in Fig.~3 of the main manuscript are reported in Table~\ref{Table2n}.\\


\subsubsection*{Attosecond pulse shaping: three-harmonic case}
For the phase shaping, correlation plots as a function of the delay introduced by the phase shifter $\mathrm{PS_4}$ were acquired. The delay introduced an additional phase given by $-7\omega_{UV}\tau_{s4}$ (see Eq.~(\ref{Eq-calibration})). The correlation coefficients $\rho_{7,8,9}$ were fitted using a cosine function and the maxima of the fit were used as reference points for the condition $\Delta\varphi_{7,8,9}=2m\pi$ where $m$ is an integer. Using this calibration, a value of the phase difference $\Delta\varphi_{7,8,9}$ was associated with each value of delay introduced by the phase-shifter. The ambiguity between $\Delta\varphi=\varphi_0$ and $\Delta\varphi=2\pi-\varphi_0$ was resolved as outlined in the previous section.

For the amplitude shaping, different configurations for the dispersive section of the FEL and the undulator gaps were used. The reference spectrum used for the comparison is the black one shown in Fig.~\ref{Fig3}h of the main manuscript. From this position, the gap of the undulator $\mathrm{U_6}$ was changed leading to a reduction of the 7th harmonic (red spectrum). A further change in the dispersive section, which controls the gain of the FEL, leads to a reduction of also the 8th harmonic (blue spectrum). The 9th harmonic was almost unaffected by these changes. Due to the limited amount of experimental beamtime, additional combinations of undulator gaps and dispersive section were not explored. For each setting of the amplitudes, a complete phase scan using the delay introduced by the phase shifter $\mathrm{PS_4}$ was acquired. These curves were used to independently calibrate the correspondence between phase difference $\Delta\varphi_{7,8,9}$ and phase-shifter delay. The correlation plots shown in Fig.~\ref{Fig3} i, j and k of the main manuscript were acquired for the same positions of the phase shifters $\mathrm{PS_2}$ and $\mathrm{PS_4}$.\\
\subsubsection*{Attosecond pulse shaping: four-harmonic case}
For the four-harmonic shaping, the relative phases $\Delta\varphi_{7,8,9}$ and $\Delta\varphi_{8,9,10}$ were changed, without any variation of the undulator gaps and dispersive section. In these conditions the amplitude of the harmonics is unaffected within the experimental error.
We acquired a complete phase scan using the delay introduced by the phase shifter $\mathrm{PS_3}$ (for a few selected positions of the phase shifter $\mathrm{PS_2}$) and, similarly, using the delay introduced by the phase shifter $\mathrm{PS_2}$ (for a few selected positions of the phase shifter $\mathrm{PS_3}$). Using these measurements, with the support of Eq.~(\ref{Eq-calibration}), we retrieved the phase differences $\Delta\varphi_{7,8,9}$ and $\Delta\varphi_{8,9,10}$ introduced by a generic combination of the delays $\tau_{s2}$ and $\tau_{s3}$.\\

\subsubsection*{Attosecond waveform retrieval}
Using Eq.~(\ref{Eq-S16}), the intensity profile for the three-harmonic case can be expressed as:
\begin{eqnarray}\label{Eq-S18}
  &I(t)&\propto \frac{1}{2}\left|\sum_{q=7}^{9}\tilde{E}_{q}(t)\right|^2\approx\sum_{q=7}^{9}F_{q}^2f^2(t)+2F_7F_8f^2(t)\cos\left[\omega_{UV} t+\eta-\frac{1}{2}\Delta\varphi_{7,8,9}\right]+\nonumber\\
   &+&2F_8F_9f^2(t)\cos\left[\omega_{UV} t+\eta+\frac{1}{2}\Delta\varphi_{7,8,9}\right]+2F_7F_9f^2(t)\cos\left[2\omega_{UV} t+2\eta\right]
\end{eqnarray}
where $\eta=(\varphi_{9}-\varphi_7)/2$. The intensity profile is determined by the frequency $\omega_{UV}$, the phase differences $\Delta\varphi_{7,8,9}$, the amplitudes $F_q$ of the harmonics and the envelope of the single harmonic $f(t)$. The phase $\eta$ introduces only a negligible temporal shift $t\rightarrow t'=t+\eta/\omega_{UV}$ of the intensity profile of the attosecond waveforms with respect to the envelope of the single harmonic. Due to the long duration (FWHM=50 fs) of the single harmonic, we neglected the effect of the temporal delays $\tau_{sx,sy}$ on the envelope $f(t)$ in Eq.~(\ref{Eq-S18}).

The intensity profile for the four-harmonic case is given by:
\begin{eqnarray}\label{Eq-S19}
  &I(t) &\propto \frac{1}{2}\left|\sum_{q=7}^{10}\tilde{E}_{q}(t)\right|^2\approx\sum_{q=7}^{10}F_{q}^2f^2(t)+2F_7F_8f^2(t)\cos\left[\omega_{UV} t+\chi-\frac{2}{3}\Delta\varphi_{7,8,9}-\frac{1}{3}\Delta\varphi_{8,9,10}\right]+\nonumber\\
   &+&2F_8F_9f^2(t)\cos\left[\omega_{UV} t+\chi+\frac{1}{3}\Delta\varphi_{7,8,9}-\frac{1}{3}\Delta\varphi_{8,9,10}\right]+\nonumber\\
   &+&2F_9F_{10}f^2(t)\cos\left[\omega_{UV} t+\chi+\frac{1}{3}\Delta\varphi_{7,8,9}+\frac{2}{3}\Delta\varphi_{8,9,10}\right]+\nonumber\\
   &+&2F_7F_9f^2(t)\cos\left[2\omega_{UV} t+2\chi-\frac{1}{3}\Delta\varphi_{7,8,9}-\frac{2}{3}\Delta\varphi_{8,9,10}\right]+\nonumber\\
   &+&2F_8F_{10}f^2(t)\cos\left[2\omega_{UV} t+2\chi+\frac{2}{3}\Delta\varphi_{7,8,9}+\frac{1}{3}\Delta\varphi_{8,9,10}\right]+\nonumber\\
   &+&2F_7F_{10}f^2(t)\cos\left[3\omega_{UV} t+3\chi\right]
\end{eqnarray}
where $\chi=(\varphi_{10}-\varphi_7)/3$. Also in this case the phase $\chi$ introduces only a negligible temporal shift of the intensity profile of the attosecond pulses ($t\rightarrow t'=t+\chi/\omega_{UV}$).\\

The amplitudes $F_{q}$ ($q=7,8,9,10)$ of the harmonics were determined as the square root \textcolor{black}{of the integral} of the corresponding photoelectron peak without NIR field. The phase differences $\Delta\varphi_{7,8,9}$ and $\Delta\varphi_{8,9,10}$ were determined using the procedure outlined in the previous section. Equations~(\ref{Eq-S18}) and (\ref{Eq-S19}) were used to retrieve the attosecond waveforms in the three and four harmonic cases, respectively.
\subsubsection*{Single-shot timing tool for FEL experiments with sub-cycle resolution}
\begin{figure}[hb]
\centering \resizebox{1.0\hsize}{!}{\includegraphics{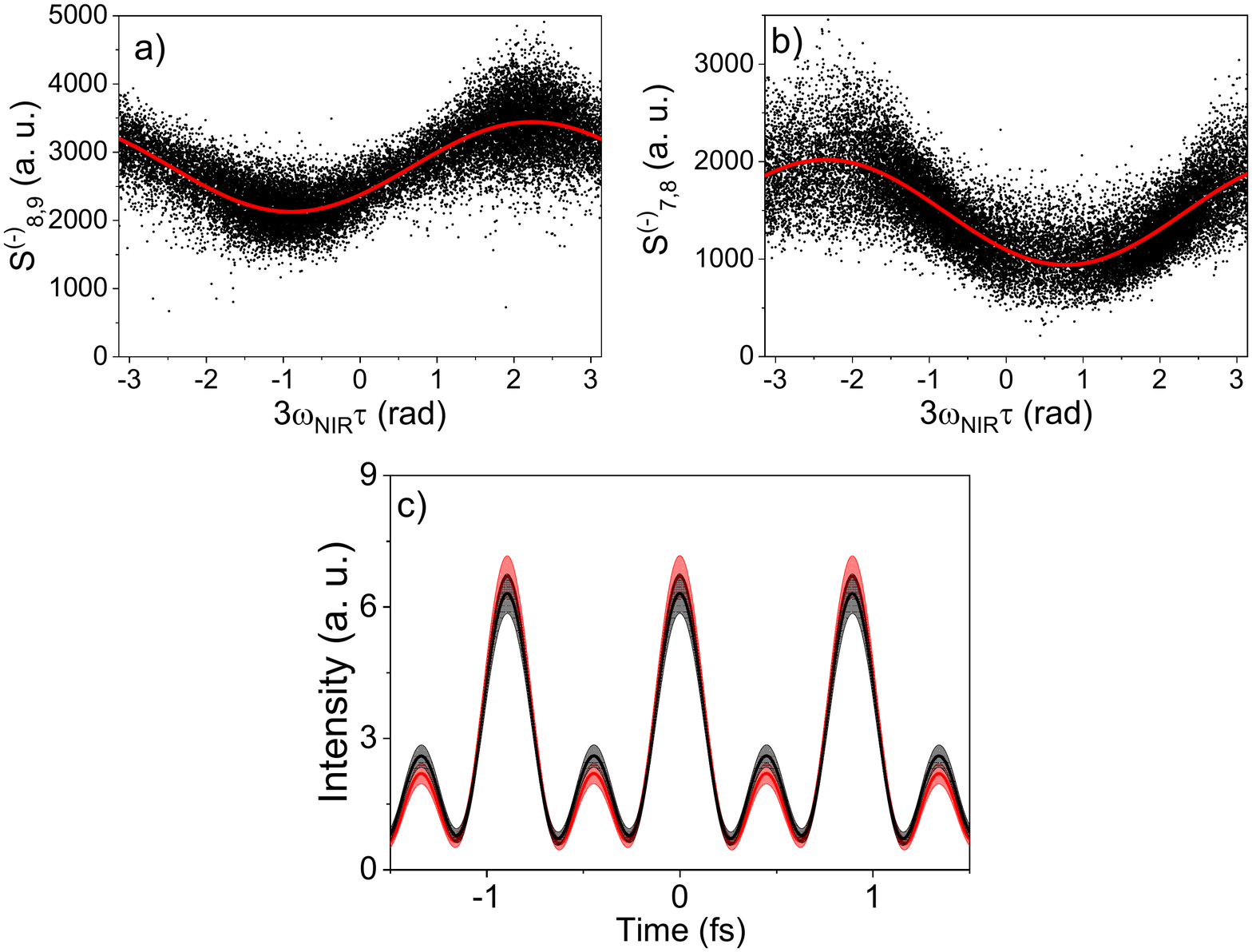}}
\caption{\textcolor{black}{Intensity of the sidebands $S^{(-)}_{8,9}$ (a) and $S^{(-)}_{7,8}$ (b) (black points) as a function of the relative phase $3\omega_{NIR}\tau$ between the attosecond pulse train and the NIR field.
The red curves show the sinusoidal fits of the distributions. Comparison of the reconstructed attosecond pulse train using the correlation parameter method $\rho_{7,8,9}$ (black curve) and the RABBIT method (red curve)
based on the phase differences extracted from the sinusoidal fits (c). The error in the reconstructions are indicated by the shaded areas.}}
\label{Fig3n_ED}
\end{figure}

\begin{figure}[hb]
\centering \resizebox{1.0\hsize}{!}{\includegraphics{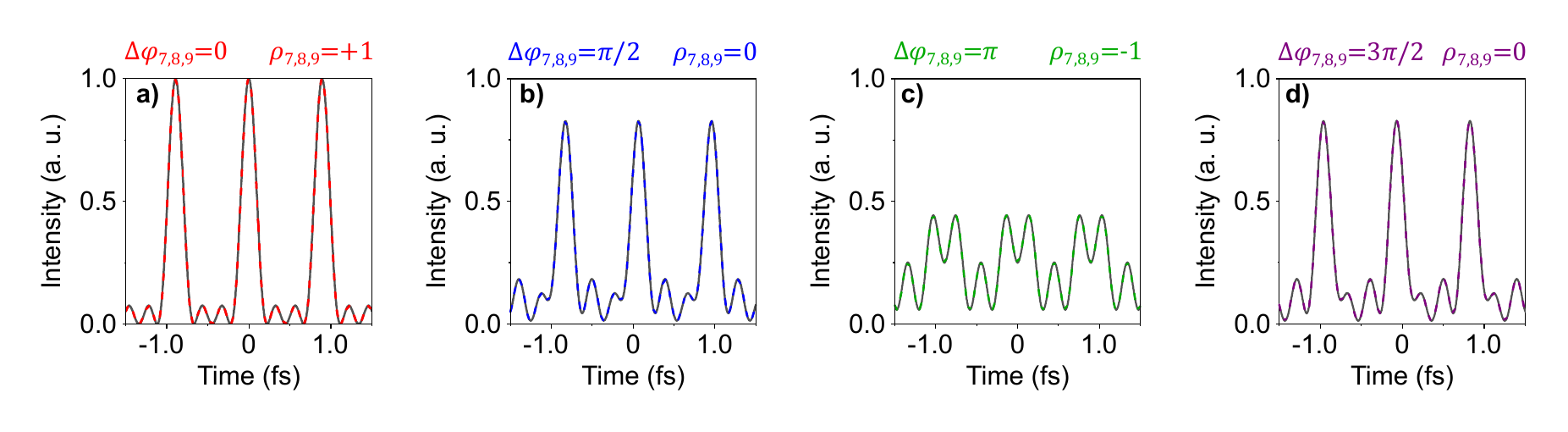}}
\centering \resizebox{0.6\hsize}{!}{\includegraphics{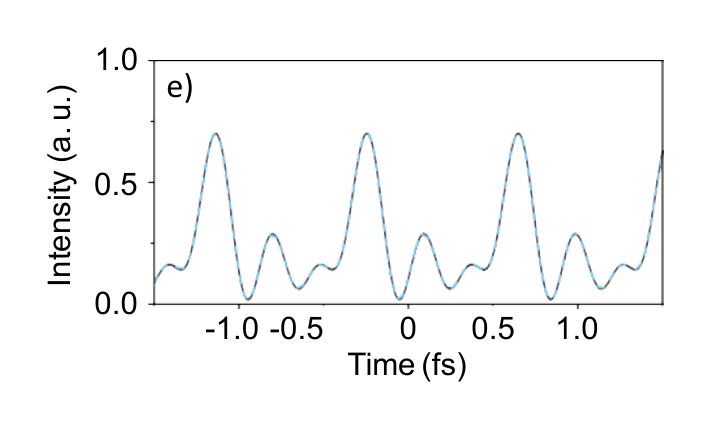}}
\caption{Reconstruction of attosecond pulses by the correlation parameter $\rho_{7,8,9}$ for multi-NIR photon transitions.
Input (black line) and reconstructed ((a) red line; (b) blue line; (c) green line; (d) violet line) intensity profiles of the attosecond train,
corresponding to Fig.~\ref{Fig1}c-f for the phase differences $\Delta\varphi_{7,8,9}$=0 (a), $\pi/2$ (b), $\pi$ (c), $3\pi/2$ (d). (e) Reconstruction of attosecond pulses by the sideband oscillations for multi-NIR photon transitions. Input (black line) and reconstructed (blue dotted line)
intensity profile of the attosecond train, corresponding to Fig.~\ref{Fig1}.  The intensity of the NIR pulse is $I_{NIR}=10^{12}~\mathrm{W/cm^2}$.
The relative phases between the harmonics are: $\varphi_{10}-\varphi_9=108^{\circ}$, $\varphi_9-\varphi_8=160^{\circ}$, $\varphi_8-\varphi_7=8^{\circ}$.}
\label{Fig8n_ED}
\end{figure}

\begin{figure}[hb]
\centering \resizebox{1.0\hsize}{!}{\includegraphics{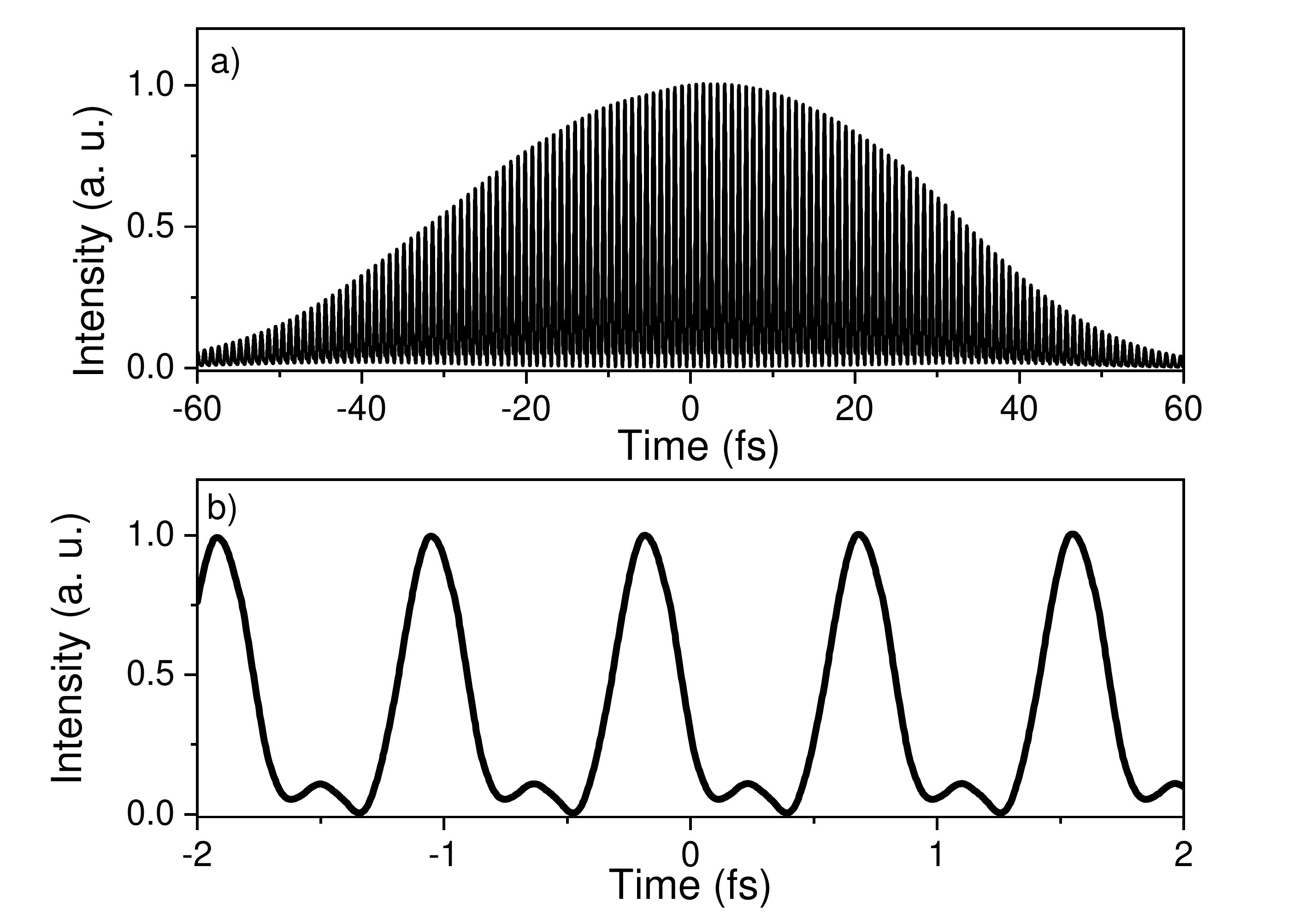}}
\caption{\textcolor{black}{Attosecond pulse train simulated using the genesis code: (a) complete temporal evolution of the train and (b) zoom on the single attosecond pulse in the train.}}
\label{Fig2n_ED}
\end{figure}

\textcolor{black}{As first application of our new technique, we present the implementation of our method as a single-shot timing tool to
retrieve the relative phase between the attosecond pulse train and the NIR field. From Eqs.~(\ref{Eq-S8},~\ref{Eq-S9},~\ref{Eq-S10}), and~(\ref{Eq-S18}) of the SI it can be easily derived that
the relative phase $3\omega_{NIR}\tau$ is related to the ratio of the single-shot experimental point
in the ($P_{7,8}, P_{8,9}$) plane by the relation:
\begin{equation}\label{delay_retrieval}
3\omega_{NIR}\tau=\arctan\left(\frac{\alpha_{8,9}\cos(\Delta\varphi_{7,8,9}/2)-\alpha_{7,8}\tan\theta\cos(\Delta\varphi_{7,8,9}/2)}{\alpha_{8,9}\sin(\Delta\varphi_{7,8,9}/2)+\alpha_{7,8}\tan\theta\sin(\Delta\varphi_{7,8,9}/2)}\right)-\eta,
\end{equation}
where $\tan\theta=P_{8,9}/P_{7,8}$.
Under the approximation $\alpha_{7,8}=\alpha_{8,9}$ and the assumption $\Delta\varphi_{7,8,9}=\pm\pi/2$ we easily obtain that:
\begin{equation}\label{delay_retrieval_special}
3\omega_{NIR}\tau= \mp (\theta - \pi/4) - \eta,
\end{equation}
and the relative phase is linearly mapped onto the angle $\theta$ in the ($P_{7,8}, P_{8,9}$) plane.
As an example, we consider the measurement shown in Fig.~\ref{Fig3}c,f of the main manuscript. Figures~\ref{Fig3n_ED} a, b report the sideband intensities as a function of the relative
phase $3\omega_{NIR}\tau$ between the XUV and NIR pulse, derived using Eq.~(\ref{delay_retrieval}). The intensities present a clear oscillation indicating the possibility to extract the relative phase from the correlation plots. 
We have then retrieved the relative phases
between the harmonics ($\varphi_9-\varphi_8$ and $\varphi_8-\varphi_7$) using a sinusoidal fit (shown in red) of the sideband oscillations. Using this information
(and the independently measured intensities of the harmonics), we have finally reconstructed the attosecond pulse trains. The comparison between the pulse train
reconstructed from the correlation parameter $\rho_{7,8,9,}$ (black line) and the RABBIT method (red line) is shown in Fig.~\ref{Fig3n_ED} c.
The agreement between the two reconstructions is a further validation of our method for the temporal characterisation of the attosecond pulse train. We estimated an error in the phase retrieval for each experimental
point of about 0.35~rad (corresponding to a delay uncertainty of 50~as in the limit of long attosecond pulse trains), mainly due to the single-shot harmonic intensity fluctuations.}

\clearpage

\pagenumbering{gobble}

\newpage

\begin{table}
  \centering
\begin{tabular}{| r | l | l | l | l | p{3cm} |}
    \hline
    &Harmonic & Photon energy (eV) & Energy $\mathrm{(\mu J)}$ & Duration $\mathrm{(fs)}$ & Intensity $\mathrm{(W/cm^2)}$ \\ \hline
    \multirow{3}{*}{\RotText{Three harm.}} & 7 & $32.88\pm0.03$  & $4.21\pm0.58$ & $50\pm5$ & $(1.1\pm0.2)\times10^{14}$ \\
    & 8 & $37.57\pm0.04$  & $5.29\pm0.58$ & $50\pm5$ & $(1.5\pm0.2)\times10^{14}$ \\
    & 9 & $42.27\pm0.04$  & $6.7\pm1.2$ & $50\pm5$ & $(1.9\pm0.4)\times10^{14}$ \\ \hline\hline
    \multirow{3}{*}{\RotText{Four harm.}} & 7 & $32.88\pm0.03$  & $1.01\pm0.12$ & $50\pm5$ & $(2.8\pm0.4)\times10^{13}$ \\
    & 8 & $37.57\pm0.04$  & $0.95\pm0.12$ & $50\pm5$ & $(2.6\pm0.4)\times10^{13}$ \\
    & 9 & $42.27\pm0.04$  & $0.67\pm0.11$ & $50\pm5$ & $(1.9\pm0.4)\times10^{13}$ \\
    & 10 & $46.96\pm0.05$ & $0.40\pm0.10$ & $50\pm5$ & $(1.1\pm0.3)\times10^{13}$ \\ \hline\hline
\end{tabular}
  \caption{\textcolor{black}{Measured harmonic order, photon energy, energy, and intensities for the three and four harmonic experiment.
  For the duration of the single harmonic we considered the values reported in ref.~\cite{PRX-Finetti-2017}.}}\label{Table1n}
\end{table}
\clearpage
\newpage

\begin{table}[hb]
  \centering
\begin{tabular}{| l | c | c | c | c | c |   p{8cm} |}
    \hline
    Panel & a & b & c & d & e    \\ \hline
    $\rho_{7,8,9}$ & $0.83\pm0.02$ & $0.74\pm0.01$ & $0.32\pm0.03$ & $-0.20\pm0.02$ & $-0.62\pm0.03$ \\ \hline
    $\Delta\varphi_{7,8,9}$ (rad) & $-0.21\pm0.12$ & $0.50\pm0.05$ & $1.21\pm0.05$ & $1.91\pm0.03$ & $2.62\pm0.05$    \\ \hline\hline
    Panel & f & g & h & i & j    \\ \hline
    $\rho_{7,8,9}$ & $-0.75\pm0.02$ & $-0.46\pm0.01$ & $0.14\pm0.03$ & $0.67\pm0.01$ & $ 0.86\pm0.01$ \\ \hline
    $\Delta\varphi_{7,8,9}$ (rad) & $3.33\pm0.20$ & $4.04\pm0.03$ & $4.74\pm0.03$ & $5.45\pm0.02$ & $6.16\pm0.02$    \\ \hline\hline
\end{tabular}
  \caption{Correlation coefficients $\rho_{7,8,9}$ and phase differences $\Delta\varphi_{7,8,9}$ for the measurements presented in Fig.~\ref{Fig2} of the main manuscript.}
  \label{Table2n}
\end{table}

\newpage

\begin{table}[h]
  \centering
\begin{tabular}{| l | l | c | r | r | r | r | r |  p{8cm} |}
    \hline
     & Figure & $F_{10}$ & $F_9$ & $F_8$ & $F_7$ & $\Delta\varphi_{8,9,10}$~(rad) & $\Delta\varphi_{7,8,9}$~(rad) \\ \hline
     \multirow{5}{*}{\RotText{Three harmonics}} & 3b,e& -- & $0.95\pm0.06$ & $1.00\pm0.03$ & $0.89\pm0.06$ & -- & $0.08\pm0.08$   \\
    & 3c,f& -- & $0.94\pm0.07$ & $1.00\pm0.03$ & $0.87\pm0.06$ & -- & $1.93\pm0.03$  \\ 
    & 3d,g& -- & $0.92\pm0.07$ & $1.00\pm0.03$ & $0.88\pm0.06$ & -- & $3.29\pm0.24$  \\ 
    & 3i,l& -- & $0.75\pm0.06$ & $1.00\pm0.04$ & $1.06\pm0.06$ & -- & $6.08\pm0.07$   \\ 
    & 3j,m& -- & $0.78\pm0.08$ & $1.02\pm0.04$ & $0.80\pm0.06$ & -- & $6.09\pm0.40$   \\ 
    & 3k,n& -- & $0.85\pm0.09$ & $0.83\pm0.05$ & $0.82\pm0.07$ & -- &$6.18\pm0.12$    \\ \hline\hline
    \multirow{2}{*}{\RotText{Four harmon.}}& 4a-c& $0.59\pm0.07$ & $0.76\pm0.06$ & $1.00\pm0.05$& $1.03\pm0.05$ & $0.20\pm0.15$ & $5.94\pm0.04$ \\
    & 4d-f& $0.58\pm0.07$ & $0.77\pm0.07$ & $1.04\pm0.05$ & $1.07\pm0.05$ & $0.20\pm0.15$ & $1.23\pm0.06$  \\ 
    & 4g-i& $0.57\pm0.08$ & $0.77\pm0.08$ & $1.03\pm0.06$ & $0.99\pm0.05$ & $2.89\pm0.08$ & $2.95\pm0.09$  \\ \hline\hline

\end{tabular}
  \caption{Amplitudes $F_{10},F_9$, $F_8$, and $F_7$ and phase differences $\Delta\varphi_{8,9,10}$ and $\Delta\varphi_{7,8,9}$ for the three- (Fig.~\ref{Fig3}) and four-harmonic cases (Fig.~\ref{Fig4}).
  For the phase (amplitude) shaping in the three-harmonic case the photoelectron spectra were rescaled to the area of the peak corresponding
  to the 8th harmonic in Fig.~\ref{Fig3}, b and e (Fig.~\ref{Fig3}, i and l).
  For the four-harmonic case the photoelectron spectra were rescaled to the area of the 8th harmonic in Fig.~\ref{Fig4}, a-c.
  \textcolor{black}{For the pulse reconstruction the measured $\Delta\varphi_{7,8,9}$ phases were corrected for the $9^{\circ}$ degree shift obtained by the
  TDSE simulations.}}\label{Table3n}
\end{table}

\newpage

\begin{table}[hb]
  \centering
\begin{tabular}{| l | c |  p{8cm} |}
    \multicolumn{2}{c}{Undulator parameters}\\
    \hline
    \hline
    U1-U2 resonant wavelength & 28.889 nm \\ \hline
    U3-U4 resonant wavelength & 32.500 nm \\ \hline
    U5-U6 resonant wavelength & 37.143 nm \\ \hline
    Undulator polarisation    & linear \\ \hline
    PS1-PS5                   & 0 rad \\ \hline
    Dispersion                & 60 $\mu$m\\ \hline
     \hline

    \multicolumn{2}{c}{Seed laser parameters}\\
    \hline
    \hline
    Wavelength & 260 nm \\ \hline
    Power  & 40 MW \\ \hline
    Pulse length (FWHM) & 110 fs\\ \hline
        \multicolumn{2}{c}{Electron beam parameters}\\
    \hline
    \hline
    Energy & 1.2 GeV \\ \hline
    Energy spread  & 110 keV \\ \hline
    Normalised emittance & 1 mm mrad \\ \hline
    Peak current & 700 A \\ \hline
    Beam size & 50-70 $\mu$m \\ \hline
\end{tabular}
  \caption{Parameters used in the Genesis code for the simulation of the generation of the train of attosecond pulses.}\label{Table4n}
\end{table}

\newpage

\clearpage

\clearpage

\clearpage

\clearpage


\clearpage

\clearpage

\clearpage

\end{document}